\documentclass[conference,compsoc]{IEEEtran}
\usepackage[T1]{fontenc}
\usepackage{amsmath}
\usepackage{amssymb}
\usepackage{array}
\usepackage{booktabs}
\usepackage{capt-of}
\usepackage{enumitem}
\usepackage{pifont}
\usepackage{graphicx}
\usepackage[caption=false,font=footnotesize]{subfig}
\usepackage{tikz}
\usetikzlibrary{positioning,fit,backgrounds}
\usepackage{cite}
\usepackage[hidelinks]{hyperref}

\newtheorem{assumption}{Assumption}
\newtheorem{lemma}{Lemma}
\newtheorem{theorem}{Theorem}
\newtheorem{proposition}{Proposition}

\newcommand{\CTA}{{\upshape\scshape CTA}}
\newcommand{\circnum}[1]{\ding{\numexpr181+#1\relax}}

\title{Temporary Authority, Permanent Effects:\\
Commit-Time Authorization for LLM Agents}

\author{Igor Santos-Grueiro}

\begin{document}
\bstctlcite{IEEEauthorlist}
\date{}
\maketitle

\begin{abstract}
An LLM agent often acts on an approval it observed earlier: it reads the
approval, prepares the request, and commits later. If the
approval was withdrawn, or covers a different effect, the service
can apply the change and report success. Seventeen fixed advisories
from three deployed agent products document failures: an approval did
not cover the executed effect, with no intervening change (12), or survived
a change in the facts it depended on (5).
We propose \emph{commit-time authorization} (\CTA), requiring at least
one complete, current justification for the exact effect at commit: a live
grant, current premises, ready dependencies, an exact-effect binding, and an
eligible path. \textsc{CommitGate} selects one complete justification. The
sink validates it or consumes an expiring, single-use derived grant atomically
with the effect. 
In a five-model service study whose prompt states the
authorization rule, timely notice prevented nearly all unauthorized
effects for GPT, Fable, and Sonnet, and fewer for Qwen and Gemma. Without a
signal, all five mostly completed tasks with unauthorized effects, even when
a read could reveal the change; Fable's counts use its provider's answered
requests. Protected sinks enforced the declared policy in LangChain approval
flows, OpenHands coding tasks, services under cloud contention, and a
transaction-runtime integration. The services and runtime refused effects
that lost authority after a positive check and preserved fresh alternatives.
After refusal, GPT, Fable, and Sonnet on OpenHands mostly published to a
separately authorized target.
Machine-checked proofs show whole-candidate
selection, trace safety for mediated atomic sinks, and exact single-use
transfer. With honest issuers, unique candidate IDs, and a premise
required by policy, every candidate \CTA{} accepts in the model carries a
live grant given at the current version; slot-by-slot checks do not ensure
this.
\end{abstract}

\section{Introduction}

An LLM agent often acts on an approval it observed earlier. While it plans,
chooses tools, or delegates, authority can change and leave the pending
action executable. Suppose an agent prepares a deployment and its only
approval is then revoked. The artifact, target, and specification stay
unchanged. A resource-version check passes, and the service delivers the
artifact against the owner's policy. The task succeeds, but the effect lacks
authority.

Approvals can also fail to cover what the agent executes. Seventeen fixed
advisories from three deployed products document effects outside an
approval's scope without an intervening change (12), or approvals that
survived changes such as a rewritten script or tightened policy (5).
The fixes check individual facts, most at execution~\cite{agentadvisories}.
LangChain and the OpenHands SDK also execute approved calls without
rereading approval or external state~\cite{langchainhitl,openhands}
(\S\ref{sec:endpoint-authority}). We also identified
authorization issues and reported them privately.
These cases pose one question: does current evidence
authorize the exact effect it is about to apply?

We propose \emph{commit-time authorization} (\CTA): at least one complete,
current justification must authorize the exact effect at its atomic commit
point. A justification contains a live grant, current \emph{premises}
(facts on which authorization depends), completed dependencies, a binding
between the executed request and the licensed effect, and permission to use the chosen
execution path. A stale justification does not remove authority supplied by
a valid alternative. Conversely, combining current claims from different,
invalid justifications cannot create an approval that no owner gave
(\S\ref{sec:model}).

These conditions often belong to different owners. An approval service
grants permission, a resource service tracks versions, and an orchestrator
records whether a worker or branch may proceed. Each owner can supply a
current fact without supplying a complete justification for the effect.
The service must establish that the facts belong together and still
authorize the operation it will perform (\S\ref{sec:model}).

Established mechanisms scope grants and
lifetimes~\cite{flask,stalesafe,authorizationcontinuity}, control tools and
untrusted data~\cite{progent,cameldefense}, coordinate
release~\cite{mindthegap,atomicityagents,atomix,cordon,masugate}, and bind
approvals or recheck authority when an effect executes
\cite{loopjacking,continuity,aidguard,bscr}. Integrating these mechanisms
requires the application to identify the exact effect, the facts on which
its approval depends, and the owners trusted to attest those facts. \CTA{}
states which claims must hold together and where the system must enforce
that decision. It also determines when another complete justification
remains sufficient after one loses validity (\S\ref{sec:related-work}).

\textsc{CommitGate} is an admission monitor placed before a protected
\emph{sink}, the component that makes the effect durable. A trusted adapter
obtains signed claims from their owners; the monitor authenticates them and
selects one complete justification. The sink then
validates authority atomically with the effect. When approval and resource
state have different owners, the approval service issues an expiring,
single-use reservation for the exact effect, and the sink consumes this
derived grant with its write (\S\ref{sec:defenses}).

A native sink can perform all of these checks under the same contract.
\textsc{CommitGate} supplies shared selection logic for services with
different policies: their adapters provide the required evidence and their
sinks apply the decision atomically. This division lets the evaluator
remain the same when the effect, evidence owners, or sink operation changes
(\S\ref{sec:boundary-enforcement}).

Machine-checked proofs in Lean establish whole-candidate selection, safety over traces of
mediated atomic commits, and single-use transfer bound to the exact effect.
With honest issuers, unique candidate identifiers, and
a premise required by policy, an accepted candidate carries a live grant
given at the current version; checking each slot separately does not ensure
this (\S\ref{sec:model}, \S\ref{sec:machine-checked-results}).

The behavioral studies follow agents through tasks whose prompt states the
authorization rule. Eighteen service tasks vary which owner's fact changes
and whether the agent receives a notice or can query status. Without a
signal, all five models mostly complete with unauthorized effects,
including changes their read tools reveal; Fable's counts use answered
requests. Timely notices prevent nearly all such effects for GPT, Fable,
and Sonnet, and fewer for Qwen and Gemma (\S\ref{sec:agent-tasks}).

The OpenHands tasks follow bug fixes through editing, testing, commit
creation, and publication~\cite{openhands}. Resource-only publication applies
all twelve revoked production writes from GPT, Fable, and Sonnet, whose
code passes the tests. Protected publication applies none. After eleven of
twelve silent protected revocations, these agents publish to separately
authorized staging. The release instructions name that fallback, so the
task can continue through an effect with its own grant
(\S\ref{sec:agent-tasks}).

The enforcement studies check whether services apply the declared policy
even when the agent submits an unauthorized effect. Protected sinks refuse
effects that lose authority after a positive check and preserve valid
alternatives. In LangChain, the adapter refuses 28 writes in 23 trajectories
whose paired native paths apply them without authority; the model has no
turn between the owner change and execution. Native sinks enforcing the
same contract are equally safe. One evaluator serves single-service
systems, reservations, Git, and Atomix
(\S\ref{sec:boundary-enforcement}). Compared with a direct authorizer,
\textsc{CommitGate} adds 0.17--0.63~ms to serial admission in mean p50 and
less than 1~ms on average to a complete 47--57~ms path on a desktop and cloud
VMs (\S\ref{sec:transfer-cost}).

We make three contributions:
\begin{itemize}
\item \textbf{Authorization contract.} \CTA{} extends the requirement for
one complete lease~\cite{authorizationcontinuity} with
premise, dependency, and eligibility claims from their owners, and requires
one complete, coherent justification at the atomic sink
(\S\ref{sec:model}).
\item \textbf{Authorization measurement.} We ground the problem in
advisories and frameworks, report authorization beside success in
$\tau$-bench, OpenHands, and LangGraph, and measure five models' responses
to notices and status tools
(\S\ref{sec:endpoint-authority}, \S\ref{sec:agent-tasks}).
\item \textbf{Enforcement and transfer.} We integrate \textsc{CommitGate}
with unmodified LangChain, reuse its evaluator across sinks, measure
cross-service authority transfer, and prove the model's guarantees
(\S\ref{sec:defenses}, \S\ref{sec:boundary-enforcement},
\S\ref{sec:transfer-cost}).
\end{itemize}

\section{Problem and Threat Model}
\label{sec:problem}

\begin{figure}[t!]
\centering
\resizebox{0.93\columnwidth}{!}{%
\begin{tikzpicture}[
    box/.style={draw, rounded corners=2pt, align=center, text width=2.05cm, minimum height=0.82cm, inner sep=3pt, font=\sffamily\footnotesize},
    observed/.style={box, fill=black!4},
    drift/.style={box, fill=black!16},
    check/.style={box, text width=2.40cm},
    bad/.style={box, text width=2.40cm, very thick, fill=black!28},
    note/.style={align=center, font=\sffamily\scriptsize, text width=2.65cm}
]
\node[observed] (observe) at (0,0) {artifact digest $d$\\approval $a_3$};
\node[observed] (derive) at (2.75,0) {prepare\\deployment};
\node[drift] (revoke) at (5.50,0) {approval revoked\\artifact stays $d$};
\node[check] (check) at (5.50,-1.42) {resource check\\still passes};
\node[bad] (commit) at (5.50,-2.84) {deployment applied\\with revoked grant};

\draw[->, thick] (observe) -- (derive);
\draw[->, thick] (derive) -- (revoke);
\draw[->, thick] (revoke) -- (check);
\draw[->, thick] (check) -- (commit);

\node[note, above=0.16cm of revoke] {target and specification\\unchanged};
\node[note, below=0.15cm of commit] {endpoint succeeds\\\CTA{} does not hold};
\end{tikzpicture}}
\caption{Revocation with an unchanged resource. The artifact,
target, and specification remain unchanged, so the resource check succeeds
after revocation. The deployment succeeds without a live grant.}
\label{fig:authorization-drift-concept}
\end{figure}

\begin{figure*}[t!]
\centering
\begin{tikzpicture}[
  box/.style={draw, rounded corners=2pt, align=center, minimum height=0.72cm, inner sep=3pt, font=\sffamily\scriptsize},
  untrusted/.style={box, fill=black!8, minimum width=2.45cm},
  trusted/.style={box, fill=black!18, minimum width=2.65cm},
  sink/.style={box, double, fill=white, minimum width=2.9cm},
  note/.style={font=\sffamily\tiny, align=center}
]
\node[untrusted] (agent) {agent-derived state\\effect + identifiers};
\node[trusted, right=0.85cm of agent] (adapter) {typed sink adapter\\policy + certificates};
\node[trusted, right=0.85cm of adapter] (monitor) {\textsc{CommitGate}\\authenticate + select};
\node[sink, right=0.85cm of monitor] (sinknode) {protected sink\\validate + mutate atomically};
\node[trusted, below=0.26cm of sinknode] (state) {authoritative state\\version, path, use record};
\node[trusted, above=0.26cm of adapter] (issuers) {issuers trusted by slot\\claims or derived grant};
\draw[->, thick] (agent) -- (adapter);
\draw[->, thick] (issuers) -- (adapter);
\draw[->, thick] (adapter) -- (monitor);
\draw[->, thick] (monitor) -- (sinknode);
\draw[<->, thick] (sinknode) -- (state);
\begin{scope}[on background layer]
\node[draw,dashed,inner sep=5pt,fit=(issuers)(adapter)(monitor)(sinknode)(state)] (tcb) {};
\end{scope}
\node[note,below=0.04cm of tcb] {trusted computing base};
\end{tikzpicture}
\caption{Enforcement path. Model output cannot self-issue evidence.
\textsc{CommitGate} authenticates and selects a candidate; the sink
validates current authority in the operation that makes the effect durable.}
\label{fig:commitgate-architecture}
\end{figure*}

\subsection{Effects, Evidence, and State Owners}

A \emph{durable effect} changes externally consequential state: it updates
an account, sends a message, writes a ticket or repository, deploys an
artifact, or changes shared agent state. Its \emph{commit boundary} is
the last point where the system can withhold it. A \emph{protected sink}
checks authorization atomically with that transition. In
Figure~\ref{fig:authorization-drift-concept}, the approval service revokes
$a_3$, the deployment's only grant. The artifact owner keeps the same digest,
target, and specification, so the resource-version check still passes. The
two owners' facts change independently.

An effect can lack authority even when every presented claim is current.
An \emph{effect-binding mismatch} occurs when a request targets ticket~99
but its only complete justification authorizes ticket~42. The grant can
remain live and both tickets unchanged. Its target binding still excludes
ticket~99. Twelve advisories in \S\ref{sec:endpoint-authority} exhibit this
mismatch between an approval's
scope and the executed effect.

\emph{Evidence pooling} combines claims from different justifications.
Suppose approval~3 was given at ticket version~7. The ticket advances to
version~8, approval~4 is given for that version and later revoked, and
approval~3 remains live. A check that accepts a live grant and a current
premise independently can combine approval~3's grant with approval~4's
version-8 premise. Both claims are authentic and current, yet no live
approval covers the change at version~8. \CTA{} rejects the combination
because its claims name different candidates. Approval~3 has a stale
premise; approval~4 has a revoked grant
(Propositions~\ref{prop:boundary-insufficiency}
and~\ref{prop:versioned-grants}).

Authority can also be lost after a valid observation. Committing after
\emph{approval loss}, as in Figure~\ref{fig:authorization-drift-concept},
or after an approved script is rewritten or a symlink retargeted, is a
time-of-check to time-of-use (TOCTOU) failure~\cite{tsafrir08tocttou}.
\CTA{} requires a complete, current justification in each case; a valid
alternative remains sufficient (\S\ref{sec:model}).

\subsection{Threat Model}

\noindent\textbf{Adversary and concurrent change.}
The victim is the user or service owner whose policy constrains the effect.
The adversary controls page, tool, or worker content that can induce the
agent to propose another effect or present its candidates. It may delay or
reorder a workflow after the agent observes a valid justification. A user,
service, or concurrent process may revoke approval, change a resource,
retarget an object, or cancel a branch. The adversary benefits if the action
remains executable. Enforcement must also cover benign concurrency.
Pooling requires no forgery: a component can assemble current claims slot
by slot, including claims carried through channels the agent influences.

\noindent\textbf{Trusted components.}
The trusted computing base comprises the sink, \textsc{CommitGate}, the
adapter policy, and the authoritative issuers of grants, versions,
identities, barriers, and branch epochs. The agent, prompts, page text, tool
responses, and worker payloads are untrusted. Trusted adapters or services
obtain claims directly from their issuers. The agent can carry candidate
identifiers but cannot obtain or alter claims.
Issuers must use unique candidate identifiers and issue claims honestly:
each signs at most one claim per candidate and slot, and each candidate's
premise names the version current when its grant was given
(Proposition~\ref{prop:versioned-grants}). Candidate assemblers must
obtain that version from the grant record.
Signatures authenticate claims. The policy must declare every relevant
fact, every protected effect must pass through the sink, and the sink must
enforce the decision atomically.

\noindent\textbf{Security goal.}
A protected effect requires at least one complete, valid candidate at the
sink's atomic commit point; missing or unavailable evidence can cause a safe
abort. With separate approval and resource owners, a bounded reservation
supplies the grant (\S\ref{sec:authority-transfer}).

\noindent\textbf{Out of scope.} The guarantee excludes:
\begin{itemize}
\item compromised trusted components, including issuers that sign dishonestly
or reuse candidate identifiers;
\item facts the policy omits and effects that bypass the sink;
\item completion, recovery, or selection among authorized effects; delay,
revocation, contention, or issuer outages may abort effects or let
reservations expire unused;
\item information already disclosed; disclosure at dispatch requires
mediation at that boundary~\cite{ghosttoolcalls}.
\end{itemize}

\begin{table}[t]
\centering
\caption{Typed evidence at a protected commit boundary.}
\label{tab:commit-boundary-observables}
\footnotesize
\setlength{\tabcolsep}{3pt}
\begin{tabular}{@{}>{\raggedright\arraybackslash}p{0.19\columnwidth}>{\raggedright\arraybackslash}p{0.32\columnwidth}>{\raggedright\arraybackslash}p{0.39\columnwidth}@{}}
\toprule
Slot & Examples & Boundary test \\
\midrule
Grant & approval, consent, capability, policy release & issuer trusted for this slot; live and names the exact effect \\
\addlinespace[2pt]
Premise & element identity, version tag, expiry, worker version & authoritative value matches the claim \\
\addlinespace[2pt]
Dependency & barrier, predecessor, callback, worker completion & dependency is resolved before commit \\
\addlinespace[2pt]
Binding & effect id, target, operation, payload digest & sink request equals the licensed effect \\
\addlinespace[2pt]
Eligibility & approval epoch, branch capability, live path & path is not revoked, canceled, or superseded \\
\bottomrule
\end{tabular}
\end{table}

\section{Commit-Time Authorization (\CTA)}
\label{sec:model}
\subsection{Evidence for One Exact Effect}

The protected sink (Figure~\ref{fig:commitgate-architecture}) evaluates a
request in configuration $x=(\sigma,c)$, where $\sigma$
is the authoritative state and
$c=(\mathit{id},\mathit{target},\mathit{op},\mathit{digest})$ is the exact
effect presented for commit. A sink policy $\Pi$ declares which evidence the
class of effect requires. An \emph{evidence slot} holds one type of claim
and has a set of trusted issuers, given by the relation $I_{\Pi}$. Every valid policy
requires an authorization grant and an exact-effect binding, and it assigns
issuers only to declared slots.

A \emph{candidate justification} $J=(g_J,P_J,O_J,b_J,e_J)$, or
\emph{candidate}, contains an authorization grant, operational premises,
dependency obligations, an exact-effect binding, and path eligibility
(Table~\ref{tab:commit-boundary-observables}). Each
component is a typed claim or bundle in one policy slot. A
\emph{certificate} is the signed object carrying these claims.
For example, candidate $J_3$ authorizes \texttt{update-ticket-42}. Its grant
permits \texttt{set\_status} at approval epoch~3, and its premise names
ticket version~7, at which approval~3 was given. Its binding names the effect
id, target, operation, and digest. Its dependency requires finalized
approval; with no separate path condition, $e_{J_3}=\top$. Each service
signs its claim together with the identifier of the candidate for which it
was issued.

A policy may use $\top$ only for an optional premise, dependency, or
eligibility condition.
$\mathsf{Shape}_{\Pi}(J)$ requires every declared slot exactly once,
$\mathsf{Trusted}_{\Pi}(J)$ checks the declared type and authenticates each
claim against $I_{\Pi}$, and $\mathsf{Coherent}(J)$ requires every claim to
name $J$. A candidate is valid when all these conditions hold:
\begin{equation}
\begin{aligned}
\mathsf{Valid}_{\Pi}(x,J) \equiv {}&
 \mathsf{Shape}_{\Pi}(J)
 \land \mathsf{Trusted}_{\Pi}(J)
 \land \mathsf{Coherent}(J) \\
&\land\ \mathsf{Live}_{\sigma}(g_J,c)
 \land \mathsf{Current}_{\sigma}(P_J) \\
&\land\ \mathsf{Ready}_{\sigma}(O_J) \\
&\land\ \mathsf{Bound}(b_J,c)
 \land \mathsf{Eligible}_{\sigma}(e_J).
\end{aligned}
\label{eq:certificate-validity}
\end{equation}
For a set $\mathcal C$ of alternative candidate justifications carried by the
pending effect,
\begin{equation}
\mathsf{CTA}_{\Pi}(x,\mathcal C)
 \equiv \exists J\in\mathcal C.\ \mathsf{Valid}_{\Pi}(x,J).
\label{eq:cta}
\end{equation}
One complete valid candidate suffices, even when another has become stale.
As with one complete lease~\cite{authorizationcontinuity}, the same candidate
must satisfy every declared slot ($\exists J\,\forall$slot). Checking that
each slot has some current claim permits the pooling of
\S\ref{sec:problem}. A \CTA{} candidate combines claims from different
owners, so $\mathsf{Coherent}$ requires each claim to name its candidate,
following the principle that a protocol message names its
context~\cite{abadineedham96}. This ties the premise, dependency, and path
to one grant without requiring the grant issuer to attest every fact.
A grant can instead name the resource version directly, as in the workload
of \S\ref{sec:transfer-cost}.

\noindent\textbf{Authoritative predicates.}
Trusted components resolve liveness, currency, readiness, and eligibility
from $\sigma$ at commit. The grant issuer attests which effect a grant
covers: a grant is live for $c$ only if it names the whole effect. The
binding issuer attests that the request is $c$, and the sink ensures that
$c$ names the object it mutates. A standing rule such as Allow Always meets
the grant clause when its issuer derives a claim for each exact effect from
a recorded scope covering the operation's full input. Facts stated by the
request, such as its target, arguments, or approved object, belong to the
binding. External facts assumed by the decision, such as file content or a
path's resolution, are premises.

Let $\mathsf{Select}_{\Pi}(x,\mathcal C)$ return the first valid candidate in
the order the trusted adapter chooses, or $\bot$ when none exists; the order
does not affect safety but determines which candidate a commit records and,
for a consumable grant, consumes.

\begin{lemma}[Whole-candidate selection]
\label{lem:whole-certificate-selection}
If $\mathsf{Select}_{\Pi}(x,\mathcal C)=J$, then
$J\in\mathcal C$, $\mathsf{Valid}_{\Pi}(x,J)$, and every claim of $J$
names it.
\end{lemma}

The proof searches the candidate list for a witness to Equation~\ref{eq:cta}.
When all claims were issued and candidate identifiers are unique, Lean also
proves that every claim of $J$ comes from $J$'s own issuance.

\subsection{Temporal Authority Divergence}

Let $x_d=(\sigma_d,c_d)$ be the configuration when the request is derived and
$x_c=(\sigma_c,c_c)$ the configuration at commit. A justification that was
valid at derivation can lose validity before commit. We call this
\emph{temporal authority divergence}:
\begin{equation}
\begin{aligned}
\mathsf{TDiv}_{\Pi}(x_d,x_c,\mathcal C)
 &\equiv \exists J\in\mathcal C.\ \mathsf{Valid}_{\Pi}(x_d,J)\\
 &\phantom{\equiv \exists J\in\mathcal C.}\land
 \neg\mathsf{Valid}_{\Pi}(x_c,J).
\end{aligned}
\label{eq:temporal-authority-divergence}
\end{equation}
The same justification can lose validity because state changes
($\sigma_d\ne\sigma_c$), the proposed effect changes ($c_d\ne c_c$), or both.
If candidate A becomes stale while B still authorizes the same effect,
divergence occurs and \CTA{} holds. Loss of all commit-time authorization is
stronger:
\begin{equation}
\mathsf{LostCTA}_{\Pi}(x_d,x_c,\mathcal C)
 \equiv \mathsf{CTA}_{\Pi}(x_d,\mathcal C)
 \land \neg\mathsf{CTA}_{\Pi}(x_c,\mathcal C).
\label{eq:lost-cta}
\end{equation}

\begin{lemma}[\CTA{} loss entails temporal divergence]
\label{lem:authorization-loss-temporal-divergence}
$\mathsf{LostCTA}_{\Pi}(x_d,x_c,\mathcal C)$ implies
$\mathsf{TDiv}_{\Pi}(x_d,x_c,\mathcal C)$; the converse does not hold.
\end{lemma}

Lean checks the implication and a surviving alternative that disproves the
converse. An effect that never had a valid justification can also violate
\CTA{} without divergence.

\begin{table*}[t]
\centering
\caption{Policy manifest: ownership and enforcement of each boundary field.}
\label{tab:policy-manifest}
\footnotesize
\setlength{\tabcolsep}{4pt}
\begin{tabular}{@{}>{\raggedright\arraybackslash}p{0.19\linewidth}>{\raggedright\arraybackslash}p{0.25\linewidth}>{\raggedright\arraybackslash}p{0.50\linewidth}@{}}
\toprule
Contract item & Authoritative owner & Enforcement consequence \\
\midrule
Grant & user or policy service & authenticate the grant claim and check that it is live and names the whole effect \\
Premises & resource owner & resolve current versions, document identity, or worker epochs \\
Binding & runtime, resource owner, or orchestrator & compare effect id, target, operation, and digest with the effect \\
Dependencies, eligibility & approval service or orchestrator & resolve worker completion and live branch capability \\
Atomic operation & resource service & recheck candidates or consume a derived grant in the write's lock or conditional request \\
Audit record & monitor and sink & record certificate ids, monitor and sink decisions, and committed target \\
\bottomrule
\end{tabular}
\end{table*}

\begin{table*}[t]
\centering
\caption{Adapter obligations and atomic operations.}
\label{tab:adapter-integrations}
\footnotesize
\setlength{\tabcolsep}{4pt}
\begin{tabular}{@{}>{\raggedright\arraybackslash}p{0.10\linewidth}>{\raggedright\arraybackslash}p{0.23\linewidth}>{\raggedright\arraybackslash}p{0.25\linewidth}>{\raggedright\arraybackslash}p{0.33\linewidth}@{}}
\toprule
System & Authoritative issuers & Exact binding & Atomic sink operation \\
\midrule
Browser & approval service; mediated browser runtime & document epoch, node/entity identity, operation, payload digest & validate identity and mutate the bound entity under the service lock \\
Tool/API & approval service; ticket service & ticket id, version, approval epoch, operation, payload digest & one owner: validate claims; separate owners: consume a derived reservation with the write \\
Multi-agent & policy service; orchestrator & worker and branch epochs, target, operation, artifact digest & validate branch capability and worker readiness while consuming the effect \\
\bottomrule
\end{tabular}
\end{table*}

\subsection{What Earlier Checks Leave Unresolved}

Let $c_t$ be the effect \texttt{update-ticket-42} of $J_3$, let
$x_0=(\sigma_0,c_t)$ be the configuration certified by $J_3$, and let
$x_1=(\sigma_1,c_t)$ differ only because approval~3 has been superseded; the
ticket remains at version~7. Let
$c_{99}$ instead target ticket~99 while the authoritative state remains
$\sigma_0$, and let $\mathcal C_3=\{J_3\}$. Last, let $x_p=(\sigma_p,c_t)$ hold
ticket version~8, live approval~3, and revoked approval~4; $J_4'$ is issued
for approval~4 at version~8, $\mathcal C_p=\{J_3,J_4'\}$, and $J_p$ is $J_3$
with $J_4'$'s premise in place of its own.

\begin{proposition}[Boundary insufficiency]
\label{prop:boundary-insufficiency}
The executable model checks
\[
\begin{aligned}
\mathsf{IfMatch}_{\sigma_1}(c_t,7)&=\mathsf{true}
 \land \neg\mathsf{CTA}_{\Pi}(x_1,\mathcal C_3),\\
\mathsf{Select}_{\Pi}(x_0,\mathcal C_3)&=J_3,\\
\mathsf{Select}_{\Pi}(x_1,\mathcal C_3)&=\bot,\\
&\mathsf{TDiv}_{\Pi}((\sigma_0,c_t),(\sigma_0,c_{99}),\mathcal C_3),\\
\mathsf{Select}_{\Pi}(x_p,\mathcal C_p)&=\mathsf{Select}_{\Pi}(x_p,\{J_p\})=\bot.
\end{aligned}
\]
\end{proposition}

The resource check passes after approval is superseded (line~1), while
selection changes from $J_3$ to refusal (lines~2--3). Changing only the
target to ticket~99 also invalidates $J_3$ (line~4). In line~5, $J_3$ has a
stale premise and $J_4'$ a revoked grant. The pooled candidate $J_p$ has
authentic, current claims, and its grant and binding name $c_t$; only
$\mathsf{Coherent}$ rejects it. Resource equality, prior admission, and
individually current claims therefore do not establish \CTA.

An immutable ledger $L$ records the version at which each grant was given.
The \emph{versioned}
relation $\mathsf{AuthV}_L(\sigma,c)$ holds when some grant for $c$ is live in
$\sigma$ and was given at the current version of $c$'s target. An issuance
history is \emph{honest} when each candidate's premise names the version at
which its grant was given. A \emph{slot-wise} check omits
$\mathsf{Coherent}$ and checks each slot independently. A policy is
\emph{sound} for an authorization relation when every valid candidate
satisfies that relation.

\begin{proposition}[Versioned grants]
\label{prop:versioned-grants}
In the checked model, let $\Pi$ require a premise, and let every claim of $J$
come from an honest issuance history with unique candidate identifiers. Then
(i)~$\mathsf{Valid}_{\Pi}(x,J)$ implies $\mathsf{AuthV}_L(\sigma,c)$, and
(ii)~$\Pi$ is thus sound for $\mathsf{AuthV}_L$ on issued candidates.
With approval~3 given at version~7 and approval~4 at version~8,
(iii)~$J_p$ passes the slot-wise check at $x_p$, its claims come from the
honest issuance of $J_3$ and $J_4'$ under distinct identifiers, and
$\neg\mathsf{AuthV}_L(\sigma_p,c_t)$; and (iv)~in $\sigma_p'$, which is
$\sigma_p$ with approval~4 live, $\mathsf{AuthV}_L(\sigma_p',c_t)$ holds, yet
$\mathsf{Select}_{\Pi}((\sigma_p',c_t),\{J_p\})=\bot$ and selection over
$J_p$ followed by $\mathcal C_p$ returns $J_4'$.
\end{proposition}

The slot-wise check is unsound for $\mathsf{AuthV}_L$ even when each claim
was honestly issued~(iii). Both issuance hypotheses matter: reusing an
identifier or pairing approval~3's grant with a version-8 premise produces
a valid candidate for an unauthorized effect. Coherence can reject a mixed
candidate even when the effect has authority, but selection still accepts
the live grant's own complete candidate~(iv).

\section{\textsc{CommitGate} Design}
\label{sec:defenses}

The agent proposes an effect and carries candidate identifiers through
pending actions, worker messages, and staged writes
(Figure~\ref{fig:commitgate-architecture}). A trusted adapter supplies the
policy and claims from their owners
(Table~\ref{tab:commit-boundary-observables}).

\subsection{Selection and Atomic Enforcement}

\begin{figure}[t]
\centering
\footnotesize
\setlength{\tabcolsep}{3pt}
\begin{tabular}{@{}r>{\raggedright\arraybackslash}p{0.88\columnwidth}@{}}
\toprule
\multicolumn{2}{@{}l@{}}{\textbf{Procedure} $\mathsf{Admit}_{\Pi}(c,\mathcal C)$} \\
\multicolumn{2}{@{}l@{}}{\textbf{Input:} exact effect $c$; ordered candidate set $\mathcal C$; policy $\Pi$} \\
\multicolumn{2}{@{}l@{}}{\textbf{Result:} boundary outcome or atomic sink result} \\
\midrule
1 & $m\gets(\mathcal C=\varnothing)$ \hfill\emph{// missing evidence seen} \\
2 & \textbf{for each} $J\in\mathcal C$, in candidate order \textbf{do} \\
3 & \quad \textbf{if} $\neg\mathsf{Shape}_{\Pi}(J)$ \textbf{then}
    $m\gets\mathsf{true}$; \textbf{continue} \\
4 & \quad $\hat x\gets(\mathsf{Resolve}_{\Pi}(J,c),c)$ \\
5 & \quad \textbf{if} $\mathsf{Valid}_{\Pi}(\hat x,J)$ \textbf{then} \\
6 & \qquad \textbf{return}
    $\mathsf{SinkCommit}_{\Pi}(c,\mathcal C,J)$ \\
7 & \textbf{if} $m$ \textbf{then return} $\mathsf{MISSING\_EVIDENCE}$ \\
8 & \textbf{return} $\mathsf{GUARDED\_ABORT}$ \\
\bottomrule
\end{tabular}
\caption{Admission algorithm. Lines~1--3 detect missing declared evidence,
lines~2--5 select a complete candidate under
Equation~\ref{eq:certificate-validity}, and in line~6 the sink repeats
selection over $\mathcal C$, or consumes a reservation issued from $J$,
atomically with the effect.}
\label{fig:commitgate-pseudocode}
\end{figure}

\begin{table*}[t]
\centering
\caption{Primary evaluation design. Each study has its own execution unit and
source of authorization decisions.}
\label{tab:denominator-map}
\footnotesize
\setlength{\tabcolsep}{4pt}
\begin{tabular}{@{}>{\raggedright\arraybackslash}p{0.17\linewidth}>{\raggedright\arraybackslash}p{0.40\linewidth}>{\raggedright\arraybackslash}p{0.33\linewidth}l@{}}
\toprule
Study & Executions & Authorization decision & RQ \\
\midrule
Fixed advisories & 17 advisories from three agent products & vendor advisory text at the primary source & RQ1 \\
Original $\tau$-bench tasks & 3 tasks $\times$ 5 conditions $\times$ 5 models $=$ 75 & grant state at each mutation; official scorer & RQ1 \\
Controlled agent study & 54 tasks $\times$ 5 conditions $\times$ 5 models $=$ 1,350 & validation flag, shown to the model & RQ2 \\
Service-agent study & 18 tasks $\times$ 16 conditions $\times$ 5 models $=$ 1,440 & owner records; the oracle reads no model text & RQ2 \\
OpenHands coding tasks & 2 tasks $\times$ 4 conditions $\times$ 2 repetitions $\times$ 5 models $=$ 80 & release-owner grant and atomic Git publication & RQ2 \\
Authoritative systems & 3 systems $\times$ 8 cases $\times$ 4 defenses $=$ 96 & authenticated claims and state at the atomic sink & RQ3 \\
Single-service panel & 18 executions $\times$ 5 defenses $\times$ 5 models $=$ 450 & claims and state at ticket and branch sinks & RQ3 \\
Cross-service panels & 5 matched 60-run panels $=$ 300 & ticket sink, with a separate approval service & RQ3 \\
Atomix composition & 5 conditions $\times$ 2 routes $\times$ 2 configurations $=$ 20 & atomic SQL sink and independent relational labels & RQ3 \\
LangChain approval path & scripted 25; model-driven 120, plus 25 paired native trajectories & owner logs; oracle written without \textsc{CommitGate} & RQ3 \\
Reservation workloads & desktop: 7,200 latency, 34,400 contention offers; VMs: 72 serial, 36 disjoint and 72 shared contention cells & owner transactions and client timing & RQ4 \\
\bottomrule
\end{tabular}
\end{table*}

\noindent\textbf{Selection.}
For each candidate, \textsc{CommitGate} \circnum{1}~verifies the schema,
\circnum{2}~authenticates claims, \circnum{3}~resolves current state, and
\circnum{4}~checks effect binding. It selects at
most one complete candidate. If none passes, absent declared evidence
produces a \emph{missing-evidence abort};
complete but stale, mismatched, unresolved, or ineligible evidence produces
a \emph{guarded abort}. This initial check, or \emph{preflight}, remains
provisional: state may change before the sink call (a \emph{post-preflight
race}), or a reservation may expire or be consumed.

\noindent\textbf{Atomic enforcement.}
The sink validates authority atomically with the effect
(Figure~\ref{fig:commitgate-pseudocode}, line~6). One state owner can use a
lock, compare-and-swap, or native conditional request
(Table~\ref{tab:policy-manifest}); adapters bind the object and operation
to that check (Table~\ref{tab:adapter-integrations}). Safety requires:

\begin{assumption}[Evidence integrity]
\label{asm:evidence-integrity}
Policy $\Pi$ declares every authorization-relevant slot, requires grant and
binding evidence, accepts each claim type only from an issuer trusted for
its slot, and binds the effect the sink executes; each issuer signs at
most one claim per candidate and slot.
\end{assumption}

\begin{assumption}[Mediation and atomicity]
\label{asm:mediation-atomicity}
Every protected effect passes through the sink, and validation or
derived-grant consumption is atomic with the mutation.
\end{assumption}

Write $S\xrightarrow{\mathsf{commit}(c,\mathcal C)}_{\Pi}S'$ for a protected
commit transition whose pre-state contains $x=(\sigma,c)$.

\begin{proposition}[Commit safety]
\label{prop:atomic-commit-safety}
In the checked model,
\[
S\xrightarrow{\mathsf{commit}(c,\mathcal C)}_{\Pi}S'
\quad\Longrightarrow\quad
\exists J\in\mathcal C.\ \mathsf{Valid}_{\Pi}(x,J).
\]
\end{proposition}

The sink must find a valid candidate in its pre-state; preflight alone is
insufficient.
Theorem~\ref{thm:trace-safety} extends this result to traces.

Our three authoritative sinks repeat selection under their commit lock.
The Atomix SQLite sink
revalidates the selected candidate in its write transaction. The Git sink
checks the grant ref and the target's old value in one ref transaction; it
relies on \textsc{CommitGate} for shape, authentication, coherence, and
binding, which do not depend on state. When a sink delegates these checks,
\textsc{CommitGate} is necessary for safety. A sink that performs them all
itself gains candidate ordering, diagnosis of missing or stale evidence, and
early refusal from the monitor.

\subsection{Cross-Service Authority Transfer}
\label{sec:authority-transfer}

A local write cannot keep another service's approval observation current.
An \emph{exact-effect reservation} $q$ transfers bounded authority to the
resource sink. The approval service checks the source candidate $J$ selected
by \textsc{CommitGate}, then issues $q$ as a derived grant. It binds $q$ to
$J$'s grant, policy version, exact effect $c$, resource version, expiry
$\tau$, and a fresh single-use identifier (\emph{nonce}) $n$. The sink authenticates and consumes $q$
atomically with the mutation.

The checked model gives the approval service and sink separate states and
clocks. \emph{Tied issuance} requires a premise slot and takes the
reservation's resource version from $J$'s premise and its policy version
from the source policy. Write $q=\mathsf{IssueT}(A,J,c,n,\tau)$ for issuance in
approval state $A$ with world $\sigma_A$, and
$\mathsf{Consume}(S,q,c')=S'$ for consumption in sink state $S$.

\begin{theorem}[Tied reservation]
\label{thm:reservation-transfer}
In the checked model, let $q=\mathsf{IssueT}(A,J,c,n,\tau)$. Then
(i)~$\mathsf{Valid}_{\Pi}((\sigma_A,c),J)$, and $q$ names $J$'s grant, $c$,
and $n$. If also $\mathsf{Consume}(S,q,c')=S'$, then (ii)~$c'=c$, and every
later consumption with nonce $n$ is denied; (iii)~the target and version of
$S$ are those of $J$'s premise, and $S$ accepts the source policy's version;
and (iv)~if each clock is within $\epsilon$
of a common time, consumption occurs at most $\tau+2\epsilon$ after
issuance.
\end{theorem}

Consumption satisfies \CTA{} under a derived policy: $q$ supplies the grant,
and the sink checks the premise and binding. Source clauses may have become
stale by up to $\tau+2\epsilon$, analogous to weak
stale-safety~\cite{stalesafe}. Like a lease~\cite{grayleases89}, $q$ can
survive source revocation until expiry, as a Lean trace demonstrates.
Immediate revocation requires revocable reservations or a coordinated
commit. Dependencies and path eligibility are checked at issuance; their
later loss does not revoke $q$. If the resource returns to an earlier
version number, the nonce still prevents a second use.

Our two cross-service prototypes provide this derived-policy guarantee.
Neither meets the honest-issuance hypothesis of
Proposition~\ref{prop:versioned-grants} and the issued-candidate part of
Theorem~\ref{thm:trace-safety}. In the cross-service panels, the premise
names the version observed by the agent, which the runs fix at the
approval's version. The reservation workloads
(Table~\ref{tab:denominator-map}) instead name the version in the grant.
That implementation checks validity and expiry under the sink's write lock
and persists nonce consumption with the mutation. It uses a monotonic clock
on one host and chrony-synchronized clocks across two VMs.

\subsection{Machine-Checked Model and Proofs}
\label{sec:machine-checked-results}

Lean~4~\cite{lean4} checks the model and proofs
(\S\ref{app:boundary-criterion}). Under a fixed policy $\Pi$, the
environment may revoke grants, advance
versions, cancel paths, withdraw dependencies, or replace the world between
sink steps. The protected apply step receives the presented candidates and
reruns selection against the current world in the same indivisible step,
recording the world $\sigma_r$, effect $c_r$, candidates $\mathcal C_r$, and
selected candidate.

\begin{theorem}[Trace commit safety]
\label{thm:trace-safety}
In the checked model, run a trace from a state with an empty commit log in
which no write bypasses the sink and every commit is one atomic step. Then
every commit record $r$ in the resulting log satisfies
$\mathsf{CTA}_{\Pi}((\sigma_r,c_r),\mathcal C_r)$, and its recorded candidate
is $\mathsf{Select}_{\Pi}((\sigma_r,c_r),\mathcal C_r)$. If $\Pi$ is sound for
$\mathsf{Auth}$, then $\mathsf{Auth}(\sigma_r,c_r)$ for every such $r$. If
$\Pi$ requires a premise and every presented candidate comes from an honest
issuance history with unique identifiers, then
$\mathsf{AuthV}_L(\sigma_r,c_r)$ for every such $r$.
\end{theorem}

The last part requires soundness on issued candidates
(Proposition~\ref{prop:versioned-grants}). The ticket policy is unsound over
arbitrary certificates: a reused identifier can make a pooled candidate
valid. A trace presenting $J_p$ and $\mathcal C_p$ in $\sigma_p$ meets the
theorem's hypotheses, and the sink applies nothing.

Counterexamples establish the role of the hypotheses. Applying an earlier
selection after revocation, or writing outside the sink, permits a commit
without \CTA. An incomplete policy permits a forbidden effect even at an
atomic sink. Conversely, when approval~3 is revoked and approval~4 remains
live, a sink checking only the forwarded candidate refuses, while a sink
that repeats selection commits. The selecting sink commits whenever \CTA{}
holds at that step. Applications requiring single use must declare a
consumable source grant; \CTA{} itself does not consume source
grants~\cite{consumablecredentials,agtadr0030}.

\noindent\textbf{Scope of the proofs.}
The proofs assume mediation, atomicity, policy soundness, and clock bounds.
For issued candidates, they also assume honest issuers, unique identifiers,
and claim provenance, which represents signature unforgeability. The grant
ledger is a parameter, and $\mathsf{AuthV}_L$ covers grant liveness and
version currency only. The definitions build in one claim per declared slot,
slot-scoped issuer trust, authenticity as a flag, indivisible sink steps, and
a nonce record that only grows. Outside the model lie policy completeness for
other relations, refinement by the Python services, nonce-record retention,
crash recovery, and policy changes.

\section{Evaluation}
\label{sec:experimental-setting}

The evaluation asks four questions: do approvals and task success imply
authorization (RQ1); how do agents respond to changed authority (RQ2);
which checks enforce it, and can one evaluator serve different policies
(RQ3); and what does cross-service transfer cost under expiry and contention
(RQ4)? Table~\ref{tab:denominator-map} identifies the execution units and
sources of authorization decisions.

\noindent\textbf{Outcomes and protocol.}
Each execution records task success and authorization separately. An
\emph{authorized completion} succeeds with \CTA; an \emph{unauthorized
commit} applies an effect without it. Withholding the effect yields a
\emph{safe non-completion}, or a \emph{missing-evidence abort} when declared
evidence is absent. An uninterpretable endpoint or decision is an
\emph{infrastructure failure}. Counts retain failures and malformed model
output; operational studies distinguish offers from bursts. Deterministic
cases run once per defense from fresh state and exact-effect evidence, with
interventions after request derivation or between preflight and the sink
call. Cases and expected outcomes for the LangChain, CONTINUITY, and
shared-target extensions were specified and hashed before execution.

\noindent\textbf{Systems and models.}
Every model study uses Qwen2.5-7B-Instruct (Qwen), served locally, and
Gemma~4~31B (Gemma), hosted, both at temperature zero; and GPT-5.6 Sol (GPT),
Claude Fable~5 (Fable), and Claude Sonnet~5.5 (Sonnet), accessed through
their providers' APIs at medium reasoning effort. All passed a format
diagnostic tailored to the study.

Anthropic's API refused 109 of Fable's 270 controlled executions, 42 of 90
single-service executions, and 107 of 288 service trajectories under its
cyber category, which also covers benign security work~\cite{anthropicrefusals}.
It refused none elsewhere, and no Sonnet study trajectory was refused.
Refusals varied by task and sometimes followed a protected refusal. Tables
mark Fable's answered-only counts; assigned recovery cohorts retain provider
refusals (\S\ref{app:task-library}).

\subsection{RQ1: Do Approvals and Success Imply Authorization?}
\label{sec:endpoint-authority}

\begin{table}[t]
\centering
\caption{Fixed advisories in which an approval did not cover the executed
effect; in the first five, something changed in between. Ids: first block
of the GHSA identifier. Cl.: the \CTA{} clause that the executed
effect escaped (G grant, P premise, B binding).}
\label{tab:agent-advisories}
\footnotesize
\setlength{\tabcolsep}{2pt}
\begin{tabular}{@{}l>{\raggedright\arraybackslash}p{0.295\columnwidth}>{\raggedright\arraybackslash}p{0.372\columnwidth}c@{}}
\toprule
Advisory & Approved & Executed & Cl. \\
\midrule
Cursor 24mc & MCP server entry & its command, edited later & B \\
OpenClaw 8g75 & \texttt{sh ./script.sh} & rewritten script content & P \\
OpenClaw q399 & a command & another binary on a changed \texttt{PATH} & P \\
OpenClaw mwcg & a working directory & the retargeted symlink & P \\
OpenClaw wj55 & a queued node action & it, after policy tightening & G \\
\midrule
Cursor r22h & a command shown & its hidden arguments & B \\
OpenClaw h3f9 & allowlisted command & a different argument vector & B \\
OpenClaw hwpq & a displayed command & another binary & B \\
OpenClaw 74gc & one exact command & same binary, other arguments & B \\
OpenClaw 6j27 & a wrapped command & another inner command & B \\
OpenClaw 3mq7 & a command in one directory & it, in another directory & B \\
OpenClaw 7jfq & a file transfer & it, on another node & B \\
OpenClaw 6x2m & a run on one node & the run on another node & B \\
OpenClaw 6rcp & the displayed command & arguments the display omitted & B \\
OpenClaw hjvp & a request & it, with a changed environment & B \\
Claude Code pq7j & a plugin at a pinned commit & a same-named branch tip & B \\
Claude Code q5hj & a trusted folder & another folder's hooks & B \\
\bottomrule
\end{tabular}
\end{table}

\noindent\textbf{Fixed advisories.}
Seventeen advisories meet our rules, fixed before reading
(Table~\ref{tab:agent-advisories}): the vendor advisory or CVE record must
name a fixed version, an approval,
allowlist entry, or operator policy, and a durable effect outside its scope.
The rules exclude command-parsing errors, forged or self-granted approvals,
and effects executed without an approval. Four further advisories admit a
contestable reading and are excluded. Twelve included cases require no
intervening change: the product executes an effect beyond what the approval
names, displays, or records. In five, the approval survives a change to a
script, \texttt{PATH}, symlink, configuration entry, or policy. Vendors assign
CWE-367 to four of these five and none of the twelve. Each fix checks one
fact, most at execution (\S\ref{app:task-library})~\cite{agentadvisories}.
A Jenkins Pipeline regression also rebuilt an approved script after
revocation~\cite{jenkinscve202452550}.

\noindent\textbf{Approval in agent frameworks.}
LangChain's human-in-the-loop middleware (1.4.4)
matches \texttt{approve} records to pending calls by position; OpenHands SDK
(1.47.0) executes every pending action once confirmed
\cite{langchainhitl,openhands}. Neither rereads approval policy or external
state when the tool runs. OpenAI Agents SDK (0.23.1) looks up an execution-time
decision bound to the call identifier and argument
fingerprint~\cite{openaiagentssdk}. Claude Code documents a recheck of path
resolution when opening a file~\cite{claudecodeperms}. Each addresses one
clause on a particular tool path. Codex's documentation does not state what
a one-time approval binds~\cite{codexapprovals}. During this analysis, we
identified authorization issues and reported them privately to the relevant
maintainers.

\noindent\textbf{Utility scoring.}
Three original $\tau$-bench~\cite{taubench} tasks---address change, return,
and cancellation---retain their tools, conversation logic, and official
scorer. Each model acts, with GPT simulating the customer. We add a revocable
account-grant policy outside the retail database and a status tool.
\emph{Late revocation} occurs immediately before the first mutation;
\emph{visible revocation} occurs at the first status query, which returns a
notice. Protected dispatch checks the grant atomically with the mutation.

All three unprotected late-revocation trajectories of Gemma, GPT, Fable,
and Sonnet apply unauthorized effects that the official scorer credits in
full (Table~\ref{tab:native-benchmark}). Qwen rarely completes a clean task;
it applies one revoked change, also credited. The scorer compares only the
database, so these successes receive the same credit as authorized ones.
Protected dispatch blocks every revoked effect by construction. Every model
that receives the visible notice withholds the change.

\begin{table}[t]
\centering
\caption{Original $\tau$-bench tasks: official successes/unauthorized
effects of three trajectories per condition and acting model. --: Qwen never
queried status, so the revocation never reached it.}
\label{tab:native-benchmark}
\footnotesize
\setlength{\tabcolsep}{3.2pt}
\begin{tabular}{@{}lccccc@{}}
\toprule
Condition & Qwen & Gemma & GPT & Fable & Sonnet \\
\midrule
Clean, unprotected & 0/0 & 3/0 & 3/0 & 2/0 & 3/0 \\
Clean, protected & 1/0 & 3/0 & 3/0 & 3/0 & 3/0 \\
Late revocation, unprotected & 1/1 & 3/3 & 3/3 & 3/3 & 3/3 \\
Late revocation, protected & 0/0 & 0/0 & 0/0 & 0/0 & 0/0 \\
Visible revocation, unprotected & -- & 0/0 & 0/0 & 0/0 & 0/0 \\
\bottomrule
\end{tabular}
\end{table}

\subsection{RQ2: How Do Agents Respond When Authority Changes?}
\label{sec:agent-tasks}

\begin{figure*}[t]
\centering
\begin{tikzpicture}[
  actionbox/.style={draw, rounded corners=2pt, align=left, text width=7.5cm,
    inner sep=3pt, font=\sffamily\footnotesize},
  owner/.style={actionbox, fill=black!5},
  change/.style={owner, fill=black!18},
  heading/.style={font=\sffamily\bfseries\footnotesize}
]
\node[heading] at (0,0) {OpenHands agent};
\node[heading] at (9.4,0) {Release service};
\node[actionbox] (prepare) at (0,-0.48) {Tests its fix commit; requests production preparation};
\node[owner] (read) at (9.4,-0.48) {Reads live production approval};
\node[change] (revoke) at (9.4,-1.24) {Intervention: revokes approval before the response};
\node[actionbox] (observe) at (0,-2.00) {Receives the earlier live approval observation};
\node[owner] (respond) at (9.4,-2.00) {Returns the request and earlier live observation};
\node[actionbox] (publish) at (0,-2.58) {Requests publication of the same commit to production};
\node[owner] (refuse) at (9.4,-2.58) {Admission refuses: \texttt{approval\_revoked}; no effect};
\node[actionbox] (status) at (0,-3.16) {Queries current status};
\node[owner] (current) at (9.4,-3.16) {Production approval revoked; staging grant live};
\node[actionbox] (stage) at (0,-3.74) {Prepares and publishes the commit to staging};
\node[owner] (apply) at (9.4,-3.74) {Publishes to staging under its separate grant};
\draw[->,thick] (prepare.east) -- (read.west);
\draw[->,thick] (read.south) -- (revoke.north);
\draw[->,thick] (revoke.south) -- (respond.north);
\draw[->,thick] (respond.west) -- (observe.east);
\draw[->,thick] (publish.east) -- (refuse.west);
\draw[->,thick] (status.east) -- (current.west);
\draw[->,thick] (stage.east) -- (apply.west);
\end{tikzpicture}
\caption{Recorded OpenHands recovery under silent revocation (top to bottom).
Preparation returns an approval observation read before revocation. A later
generation requests production publication, which admission refuses. The agent
then publishes to staging under a separate grant, as the release instructions
direct. Production remains unchanged.}
\label{fig:openhands-recovery}
\end{figure*}

\begin{table}[t]
\centering
\caption{Service-agent study, unprotected arm: unauthorized completions of
36 trajectories with an owner change (Wilson intervals in
\S\ref{app:task-library}); Q: status queries after it; Pairs: silent
then notice, on $n$ pairs qualifying under a rule fixed before the runs;
Clean: completions of 72. Fable: answered trajectories.}
\label{tab:service-agent}
\footnotesize
\setlength{\tabcolsep}{2.6pt}
\begin{tabular}{@{}lccccccc@{}}
\toprule
Model & Silent & Notice & Status & Q & Pairs & $n$ & Clean \\
\midrule
Qwen & 26 & 15 & 26 & 0 & 14$\to$3 & 22 & 55 \\
Gemma & 34 & 10 & 32 & 6 & 32$\to$8 & 32 & 70 \\
GPT & 25 & 1 & 2 & 34 & 17$\to$0 & 20 & 72 \\
Fable & 17/22 & 0/23 & 17/27 & 13/27 & 15$\to$0 & 19 & 46/46 \\
Sonnet & 30 & 0 & 22 & 12 & 29$\to$0 & 34 & 72 \\
\bottomrule
\end{tabular}
\end{table}

\noindent\textbf{Owner changes during a task.}
The service study tests whether agents notice changed authority and whether
a notice or status tool changes their actions
(Table~\ref{tab:service-agent}). Eighteen tasks map the controlled catalog
to durable writes on local services, with two tasks per workflow family and
kind of changed fact. A LangGraph agent reads, prepares, and writes through
standard tool calls. Its system prompt permits a change only while approval
and the conditions under which it was given still hold.

When the first preparation or write returns, the owner withdraws approval
or changes a premise version, readiness, page-element binding, or run or
branch state. The agent receives no signal (\emph{silent}), receives a
factual notice with the preparation result (\emph{notice}), or has a status
tool available throughout (\emph{status}). Each task runs clean with and
without the status tool, then under two changes with each signal, on an
unprotected path and a protected path that revalidates under its lock:
288 trajectories per model. An oracle reads only owner and sink logs to
label each applied effect at its commit instant.

\noindent\textbf{What agents commit after a change.}
Without a signal, every model mostly completes through an unauthorized
effect (Table~\ref{tab:service-agent}; Fable counts answered trajectories).
Read and wait tools reveal premise, readiness, and binding changes. Of the
14 changes these tools can reveal, unauthorized completions occur in 11
(Qwen), 13 (Gemma), 8 (GPT), 9 (Sonnet), and 6 of 11 answered cases (Fable).
The tools cannot reveal grant or path changes, after which completion is
unauthorized by construction.

The notice condition leaves at most one unauthorized completion for GPT,
Fable, and Sonnet, and reduces Gemma's from 34 to 10. Twelve of Qwen's 15
unauthorized completions issue the write together with preparation, before
a notice can arrive. A rule fixed before execution selects matched pairs
with the same task and change, identical responses before the change, and
a notice before any write reaches the service. Unauthorized completion
falls for every model on these pairs (exact McNemar tests, $p<0.004$).
The decrease also holds when grouped by task, accounting for dependence
between pairs (sign tests, $p\le0.039$; \S\ref{app:task-library}).

With a status tool, GPT queries after the change in 34/36 trajectories and
its unauthorized completions fall from 25 to 2. Sonnet's fall from 30 to
22. The remaining models query in at most half of their trajectories and
change little. Every model except Qwen completes nearly all clean
trajectories, and the protected sink refuses none of those clean writes.

The protected path has no unauthorized effect by construction: sink and
oracle apply the same rule. Of 108 protected change trajectories, 42 permit
recovery by renewing a premise, completing a dependency, or correcting a
binding. The other 66 withdraw grant or path authority without replacement.
Authorized recovery occurs in 9 (Qwen), 27 (Gemma), 30 (GPT), 16 (Sonnet),
and 29 (Fable) of the 42 assigned cases. Fable's denominator retains its
13 provider refusals (\S\ref{app:task-library}). The recovery counts measure
whether an agent finds a permitted way forward after the change; the 66
terminal cases instead require it to stop.

\begin{table}[t]
\centering
\caption{OpenHands trajectories, four per model and condition. P/S:
authorized production/staging; W: refusal without publication; U:
unauthorized production publication; X: no publication after an agent
error, the request limit, or failed verification.}
\label{tab:openhands-tasks}
\footnotesize
\setlength{\tabcolsep}{3pt}
\begin{tabular}{@{}lccccc@{}}
\toprule
Condition & Qwen & Gemma & GPT & Fable & Sonnet \\
\midrule
Clean, protected & 4X & 1P/3X & 4P & 4P & 4P \\
Silent revocation, protected & 4X & 1S/3X & 3S/1W & 4S & 4S \\
Visible revocation, protected & 4X & 4X & 4S & 4S & 4S \\
Silent revocation, resource only & 4X & 4X & 4U & 4U & 4U \\
\bottomrule
\end{tabular}
\end{table}

\noindent\textbf{OpenHands coding tasks.}
Two tasks fixing historical Requests bugs follow the agent through editing,
testing, committing, and publication (Table~\ref{tab:openhands-tasks}). The agent runs on the unmodified
OpenHands SDK. Our release service owns production and staging grants and
a bare Git repository outside the agent's sandbox. Protected publication
uses \textsc{CommitGate} and a Git ref transaction that checks the grant
revision; resource-only publication omits that grant check.

Each model executes both tasks twice under four conditions:
\circnum{1}~clean, \circnum{2}~silent revocation after approval is read,
\circnum{3}~visible revocation at the same point, and \circnum{4}~silent
revocation with only resource consistency enforced. The second
task uses OpenHands delegation. Release instructions name a status command
and direct the agent to separately approved staging if production approval
is lost. Staging is a different effect whose grant cannot authorize
production.

\noindent\textbf{Publication and recovery.}
For GPT, Fable, and Sonnet, resource-only publication applies all 12 revoked
production writes, as expected without a grant check. Protected publication
permits all 12 clean writes and no revoked write. After silent protected revocation, these
agents publish to staging in 11 trajectories and stop in one
(Figure~\ref{fig:openhands-recovery}). None queries status before the first
publication attempt. Every candidate and applied commit from these models
passes the tests, including the unauthorized publications. Qwen and Gemma
publish in 2 of 32 trajectories; 23 end in an agent error, including 22
that reach the execution limit. Their incomplete
workflows provide little evidence about recovery after refusal.

\begin{table}[t]
\centering
\caption{Controlled study: executions the model commits; each committed
flagged one is unauthorized. Fable's exclude 87 flagged and 21 control
executions its provider refused.}
\label{tab:controlled-models}
\footnotesize
\setlength{\tabcolsep}{4pt}
\begin{tabular}{@{}lcc@{}}
\toprule
Model & \shortstack{Flagged executions\\committed} & \shortstack{Clean controls\\committed} \\
\midrule
Qwen & 207/215 & 54/54 \\
Gemma & 0/215 & 25/54 \\
GPT & 4/215 & 0/54 \\
Fable & 8/128 & 26/33 \\
Sonnet & 0/215 & 22/54 \\
\bottomrule
\end{tabular}
\end{table}

\begin{table*}[t]
\centering
\caption{Defense results over three authoritative systems, the same whether
protected paths check grants by epoch or by the full predicate of
Equation~\ref{eq:certificate-validity}. A/U/S/M: authorized completion, unauthorized commit, safe
non-completion, missing-evidence abort.}
\label{tab:independent-oracle}
\footnotesize
\setlength{\tabcolsep}{3.5pt}
\begin{tabular}{@{}>{\raggedright\arraybackslash}p{0.21\linewidth}>{\raggedright\arraybackslash}p{0.13\linewidth}cccc@{}}
\toprule
Case & Expected & Unguarded & Recheck & \shortstack{Native conditional\\sink} & \shortstack{\textsc{CommitGate}\\$+$ same sink} \\
\midrule
Clean & authorized & 3A & 3A & 3A & 3A \\
Benign concurrent & authorized & 3A & 3A & 3A & 3A \\
Alternative grant & authorized & 3A & 3A & 3A & 3A \\
Stale unrelated & authorized & 3A & 3A & 3A & 3A \\
\midrule
Direct invalidation & unauthorized & 3U & 1U/2S & 3S & 3S \\
Post-preflight race & unauthorized & 3U & 3U & 3S & 3S \\
Forged claim & unauthorized & 3U & 3U & 3S & 3S \\
Missing declared slot & unauthorized & 3U & 2U/1S & 3S & 3M \\
\midrule
All 24 per defense & --- & 12A/12U & 12A/9U/3S & 12A/12S & 12A/9S/3M \\
\bottomrule
\end{tabular}
\end{table*}

\begin{table*}[t]
\centering
\caption{Cross-service results: 20 case--defense cells, each with the same
outcome in all 15 executions (three per model), also when grants are checked
by the full predicate of Equation~\ref{eq:certificate-validity}. A/U/S:
authorized completion, unauthorized commit, safe non-completion.}
\label{tab:cross-service-authority}
\footnotesize
\setlength{\tabcolsep}{4pt}
\renewcommand{\arraystretch}{1.05}
\begin{tabular}{@{}>{\raggedright\arraybackslash}p{0.20\linewidth}ccccc@{}}
\toprule
Case & Unguarded & \texttt{If-Match} & Client recheck & \shortstack{Non-atomic\\staged recheck} & \textsc{CommitGate} $+$ reservation \\
\midrule
Clean & A & A & A & A & A \\
Approval-only invalidation & U & U & S & S & S \\
Post-preflight race & U & U & U & U & S \\
Fresh alternative grant & A & A & A & A & A \\
\bottomrule
\end{tabular}
\end{table*}

\noindent\textbf{Stale records shown to the model.}
A controlled study isolates a narrower decision: commit or abort after an
explicit invalidation flag, with no opportunity to revalidate. It runs 54
synthetic browser, tool/state, and multi-agent tasks in LangGraph, once clean
and under four perturbations. The goal and payload shape stay fixed while
expiry, version advance, retargeting, reordering, a stale cached view, or
branch cancellation invalidates a required fact
(\S\ref{app:task-library}). Before its final generation, the model receives
the environment's validation result and returns \texttt{safe\_commit} or
\texttt{abort}. The environment flags 215 of 216 perturbed executions.

Qwen commits 207 flagged executions (Table~\ref{tab:controlled-models});
its endpoint success of 262/270 therefore exceeds authorized completion of
55/270. The other models rarely commit flagged records but also abort
harmless controls. GPT aborts all 54 controls, citing revalidation that the
study does not permit. These results measure decisions under the restricted
interface, leaving recovery during a task untested.

\subsection{RQ3: Which Checks Enforce Authorization?}
\label{sec:boundary-enforcement}

\noindent\textbf{Checks under one owner.}
Three deterministic systems own browser, ticket, and branch-epoch state
(Table~\ref{tab:adapter-integrations}); their issuers sign grants and
validity claims (\S\ref{app:independent-oracle}). Each runs four authorized
and four unauthorized cases from identical state under four defenses:
\circnum{1}~unguarded execution, \circnum{2}~a final-resource recheck,
\circnum{3}~a native conditional sink checking the full contract, and
\circnum{4}~\textsc{CommitGate} followed by that sink.
Expected outcomes are fixed beforehand; monitor and sink decide separately.

Across the 96 executions, both protected paths reject every unauthorized
commit and preserve all 12 authorized cases per defense
(Table~\ref{tab:independent-oracle}). The resource recheck stops two
invalidating cases and one missing-slot case, leaving nine unauthorized
commits: an ineligible branch, forged or incomplete evidence, and every
post-preflight race. The protected sinks reject those races under their
lock despite positive preflight decisions. For missing evidence,
\textsc{CommitGate} also identifies the absent slot.

\noindent\textbf{Separate approval and resource owners.}
A four-node LangGraph workflow queries approval and ticket services, asks
a model to derive tool arguments, and sends them to the ticket sink. Each
model generates the pending action three times in each of 20
case--defense cells. The environment then changes approval while ticket
version~7, target, operation, and payload remain fixed. All 300 requests
reach the sink, with the same outcome in every execution of each cell
(Table~\ref{tab:cross-service-authority}).

\texttt{If-Match}~\cite{httpsemantics} checks only the resource version
atomically, permitting both approval-only invalidation and post-preflight
races. Client and staged rechecks read both services without coordination:
they reject grants already superseded but permit those revoked afterward.
On the reservation path, \textsc{CommitGate} refuses superseded grants at
preflight, and the approval service refuses issuance after later revocation.
Every defense commits the clean case and the fresh alternative, preserving
the effect authorized by the latter.

\noindent\textbf{LangChain approval middleware.}
The adapter uses LangChain~1.4.4's documented \texttt{wrap\_tool\_call}
hook to check evidence before tool execution. A ticket service in a separate
process checks it again in its write transaction. LangChain's
human-in-the-loop middleware runs unchanged. Twelve scripted cases run once
per arm, plus one control (25 executions); an oracle labels writes from
owner logs alone. Native execution applies all 14 writes, including eight
without authority. Protected execution refuses six attempts before the tool
and three at the sink, including a withdrawal after the first check
(\S\ref{app:supplementary-checks}).

Every write covered by a current exact approval applies, including a
reviewer's edited call and a second approver's approval after the first
withdraws. Oracle and sink agree on every attempt. A constructed case pools
two approvals' current claims as in \S\ref{sec:problem}. The adapter rejects
the differing candidate identifiers; the evaluator with that check removed
applies the effect. The integration adds 187 lines, with coherence checked
by a 7-line evaluator subclass.

\noindent\textbf{Models choosing tools.}
Five models use LangChain's \texttt{create\_agent} to choose tools for five
ticket tasks under scripted approver and owner events (24 trajectories per
model). The adapter and sink mediate all 131 attempted writes, including a
tool outside the approval configuration. They apply all 62 writes with a
current exact approval and refuse the other 69. After a refusal for which
authorized recovery is possible, Qwen repeats the write and completes in
three of five cases. Each other model reads first in five of six cases and
completes in four or five. Under the declared policy, which binds approval
to the whole resource version, changing an unread field causes four Qwen
refusals and three for each other model.

A paired extension executes five readable owner-change cases natively for
each model: 25 trajectories, of which 23 reach the intervention. All 28
writes presented at those approval interrupts apply without authority;
the adapter refuses all 28 in the matched protected cases. LangChain shows
state when the model requests the write, accepts approval later, and
executes on resume without rechecking. The model has no turn between the
change and execution. After protected refusal, all 20 hosted-model
trajectories read first; they later apply 19 authorized writes and none
unauthorized.

A separate case needs no injected change. All five models group two writes
approved at version~7 into one message. The first write advances the
version, invalidating the second. Native execution applies that second
write in all five cases; protected execution refuses it in all five.
GPT and Gemma then read and write under fresh approval; Fable resends and
obtains fresh approval; Sonnet ends after reading. Qwen groups calls again
without completing the second write. Authorization holds at the sink in
these recorded continuations.

\noindent\textbf{Other frameworks and transports.}
LangGraph~1.2.10 also preserves approval resolved through
\texttt{interrupt} without expiry or withdrawal. A sink that consumes the
exact-effect grant refuses the approved append after withdrawal. In a
single-service panel of 90 LangGraph executions per model, unguarded paths
commit unauthorized effects and paths with an atomic final check commit
none (\S\ref{app:task-library}). Scripted HTTP and Chrome checks show the
same distinction. In both service panels, the model transcribes a prepared
effect before the intervention; the case and defense then determine the
outcome of this fixed action.

\noindent\textbf{Clause checks with the sink fixed.}
The same atomic operation can enforce different evidence requirements
(\S\ref{app:supplementary-checks}). Removing coherence from the LangChain
evaluator admits the pooled candidate; restoring it prevents that write
from reaching the sink. In all three services, grants naming only an epoch
let both protected paths apply another effect's grant. Signing the whole
effect into the grant, a 30-line adapter change, rejects all six such
candidates. Grant-only and binding-only candidates cause missing-evidence
aborts; a complete candidate applies, and a sibling-effect certificate is
refused.

Claims scoped to the effect's own facts preserve all nine benign cases
that service-wide counters refuse. Signing candidate identity adds 88 lines
across adapters, evaluator, and reservation path. Correct binding also
depends on the adapter: an adapter naming another effect withholds a
justified write that the native sink applies.

\noindent\textbf{One evaluator for different policies.}
The Git policy requires grant, resource-premise, and binding claims;
Atomix/SQLite requires grant and binding. Adding a review slot and issuer
invalidates the old candidate, while a complete replacement passes the
unchanged evaluator and sink (\S\ref{app:supplementary-checks}). The
227-line evaluator module is byte-identical across three single-service
systems, the reservation path, Git, and SQLite. Each single-service adapter
adds 90--101 lines to a shared base. OpenHands and Atomix run unmodified.

\noindent\textbf{Composition with Atomix.}
Atomix~\cite{atomix} coordinates effect release and delegates semantic
validation to the application. We compare resource checks without grant
validation with \textsc{CommitGate} and atomic authority validation in the
same runtime. Both configurations register resource and approval scopes,
retain Atomix's release checks, and execute callbacks in a SQLite
transaction checking request identity, resource version, and path. Five
initially authorized conditions vary grant lifetime, check timing, and
alternatives through both release routes and configurations: 20 executions,
labeled by independent SQL queries (Table~\ref{tab:atomix-authorization}).

Every callback satisfies Atomix's release dependencies. Resource checks
permit ten publications, six unauthorized. The protected composition
preserves the four authorized publications and refuses six, including two
at the sink after a positive preflight. Atomix establishes readiness for
release; the sink establishes authority in the write transaction. Twelve
further cases on the same SQLite sink give identical outcomes with a direct
authorizer and \textsc{CommitGate}: three publications, eight rejections,
and one abort.

\begin{table}[t]
\centering
\caption{Atomix publication outcomes; each entry is two identical
executions, through \texttt{commit} and \texttt{flush\_ready}.
The protected configuration adds \textsc{CommitGate} and atomic authority
validation to the same runtime and SQLite sink.}
\label{tab:atomix-authorization}
\footnotesize
\setlength{\tabcolsep}{4pt}
\renewcommand{\arraystretch}{1.15}
\begin{tabular}{@{}lll@{}}
\toprule
Condition & Resource only & Protected \\
\midrule
Current approval & Authorized & Authorized \\
Expired approval & Unauthorized & Preflight refusal \\
Revoked before preflight & Unauthorized & Preflight refusal \\
Revoked after preflight & Unauthorized & Sink refusal \\
Surviving alternative & Authorized & Authorized \\
\bottomrule
\end{tabular}
\end{table}

\begin{table}[t]
\centering
\caption{Mean p50 difference, \textsc{CommitGate} minus direct, in ms, over
six order-balanced repetitions per window. \textsc{CommitGate}'s
complete-path p50, median of repetitions: desktop 47.31--47.98, one VM
53.19--54.23, two VMs 56.34--56.93~ms.}
\label{tab:reservation-latency}
\footnotesize
\setlength{\tabcolsep}{4pt}
\begin{tabular}{@{}llrrr@{}}
\toprule
Setting & Stage & 10 s & 1 s & 100 ms \\
\midrule
Desktop & Admission & 0.26 & 0.17 & 0.24 \\
 & Consumption & 0.31 & 0.01 & $-0.03$ \\
 & Complete path & 0.94 & 0.15 & 0.13 \\
\addlinespace[2pt]
One VM & Admission & 0.63 & 0.51 & 0.51 \\
 & Consumption & 0.15 & 0.18 & 0.19 \\
 & Complete path & 0.90 & 0.92 & 0.79 \\
\addlinespace[2pt]
Two VMs & Admission & 0.36 & 0.57 & 0.43 \\
 & Consumption & $-0.12$ & 0.21 & $-0.05$ \\
 & Complete path & 0.18 & 0.88 & 0.41 \\
\bottomrule
\end{tabular}
\end{table}

\subsection{RQ4: What Does Authority Transfer Cost?}
\label{sec:transfer-cost}

\noindent\textbf{Reservation workloads.}
We measure admission, reservation issuance, and consumption
(Table~\ref{tab:reservation-latency}) in two long-lived services with
separate SQLite databases. They run over loopback
on a desktop (\S\ref{app:policy-evaluation}), then on separate cloud VMs in
one zone. A control collocates both services on a VM of the same type.
The direct authorizer and \textsc{CommitGate} receive identical evidence
and use the same policy, transfer semantics, and atomic consumption; only
the latter uses typed certificates. Client timing includes HTTP, queueing,
accounting writes, owner persistence, and admission. Cloud workloads use one
client process per worker.

\noindent\textbf{Latency.}
Six serial repetitions per window alternate arm order by block. \textsc{CommitGate} admission is slower in every window and block
across all settings. The mean p50
difference is 0.17--0.63~ms; differences by block range from 0.09 to 0.74~ms,
and one repetition reaches 1.27~ms. Identical consumption code differs by
0.15--0.19~ms between arms on one VM, indicating an arm bias of that size
in the cloud comparison. On the desktop, the
admission difference exceeds the consumption difference only at 1~s and
100~ms.

Complete-path mean differences between arms remain below 1~ms, although
a single desktop run with the direct arm first differs by 10.99~ms
(\S\ref{app:policy-evaluation}). Separating services across two VMs adds
2.4--3.0~ms.

\begin{table}[t]
\centering
\caption{Completed effects, \% of 1,600 offers per cell (100 bursts of 16).
Desktop: threaded client, direct (D) and \textsc{CommitGate} (G).
VM columns: G with one client process per worker, mean of six repetitions;
shared-target D also gives 6.25\% throughout. --: not measured.}
\label{tab:reservation-availability}
\footnotesize
\setlength{\tabcolsep}{2.8pt}
\begin{tabular}{@{}lrrrrrrrr@{}}
\toprule
& \multicolumn{2}{c}{Desktop shared} & \multicolumn{2}{c}{Desktop disjoint} & \multicolumn{2}{c}{VM disjoint} & \multicolumn{2}{c}{VM shared} \\
\cmidrule(lr){2-3}\cmidrule(lr){4-5}\cmidrule(lr){6-7}\cmidrule(l){8-9}
Window & D & G & D & G & 1 & 2 & 1 & 2 \\
\midrule
10 ms & 0 & 0 & 0 & 0 & -- & -- & -- & -- \\
100 ms & 0.13 & 0 & 0 & 0.13 & 95.50 & 96.75 & 6.25 & 6.25 \\
300 ms & -- & -- & -- & -- & 99.95 & 100 & -- & -- \\
1 s & 6.25 & 6.25 & 98.56 & 98.69 & 100 & 100 & 6.25 & 6.25 \\
10 s & 6.25 & 6.25 & 100 & 100 & -- & -- & 6.25 & 6.25 \\
\bottomrule
\end{tabular}
\end{table}

\noindent\textbf{Expiry and contention.}
Desktop contention covers six workloads: serial offers, single-worker
bursts, and four- or sixteen-worker bursts on shared or disjoint targets.
Two arms and four expiry windows give 48 configurations and 34,400 measured
offers, each making three HTTP calls. Each configuration runs once, direct arm first, so arm differences cannot
be attributed to the authorizer. With sixteen workers and a 10-s window, all disjoint offers
complete but only 6.25\% of shared-target offers do (25\% with four workers).
The first write advances the shared version; the others remain bound to
its old value. Longer windows reduce expiry losses but cannot restore an
obsolete version, and they prolong the derived grant's authority.

With sixteen workers in one client process, almost no offer completes within
100~ms, even on disjoint targets. The median client delay from issuance to
consumption is 342~ms, compared with server round trips of about 41 and
32~ms. One process per worker reduces that delay to about 21~ms. On disjoint
targets, 95.50\% of offers complete within 100~ms on one VM and 96.75\%
across two. At 300~ms, every repetition completes at least 99.69\%, with no
unauthorized effect. Clock probes estimate an inter-VM offset below
0.85~ms; the longest issuance-to-consumption interval is 361~ms, at
$\tau=1$~s.

Shared-target contention with sixteen client processes adds 72
order-balanced cells and 115,200 measured offers
(Table~\ref{tab:reservation-availability}). For every topology, arm, and
window, each of 600 bursts commits one offer and the other 9,000 offers
abort on the version check. No burst loses its winner to expiry, even at
100~ms, and no result is unknown or unauthorized. At $\tau=1$~s across two
VMs, consumption returns in 25.89~ms for version aborts and 28.96~ms for
commits, compared with 25.72 and 27.95~ms under direct authorization. The
last of sixteen responses arrives at 260.46~ms after burst release at p50
and 413.74~ms at p95 (direct: 258.75 and 377.50~ms), each the median of
six per-repetition quantiles. Every observed issuance-to-commit interval
among completed offers meets $\tau+\hat\epsilon_{\mathrm{par}}$. The maximum
is 101.69~ms at $\tau=10$~s, with an estimated inter-host timing margin
$\hat\epsilon_{\mathrm{par}}=1.63$~ms.

\section{Discussion}
\label{sec:discussion}

\begin{table*}[t]
\centering
\caption{Closest mechanisms, grouped by contract. Current effect authority:
a valid source or derived grant under its lifetime and revocation semantics;
complete evidence alternatives: one justification satisfies every condition,
without pooling. $\checkmark$: explicit obligation; $\circ$: related
mechanism or application responsibility; --: not stated.}
\label{tab:related-contracts}
\footnotesize
\setlength{\tabcolsep}{5pt}
\renewcommand{\arraystretch}{1.1}
\begin{tabular}{@{}>{\raggedright\arraybackslash}p{0.245\textwidth}>{\raggedright\arraybackslash}p{0.365\textwidth}ccc@{}}
\toprule
Contract & Mechanisms
& \shortstack{Current effect\\authority}
& \shortstack{Exact-effect\\binding}
& \shortstack{Complete evidence\\alternatives} \\
\midrule
Policy-rule alternatives & Cedar, Progent~\cite{cedar2024,progent}
& $\circ$ & $\circ$ & $\checkmark$ \\
Tool and data-flow controls & CaMeL~\cite{cameldefense}
& $\circ$ & $\circ$ & $\circ$ \\
Request-bound authority & SUDP, CapLease, SecureClaw, CXI~\cite{sudp,caplease,secureclaw,cxi}
& $\checkmark$ & $\checkmark$ & -- \\
Authorization continuity & Live effect-scoped leases~\cite{authorizationcontinuity}
& $\checkmark$ & $\circ$ & $\checkmark$ \\
Transaction settlement & Atomix~\cite{atomix}
& $\circ$ & $\circ$ & -- \\
Task validation and release & Cordon~\cite{cordon}
& $\checkmark$ & $\checkmark$ & -- \\
Policy/effect coordination & PSS / MasuGate~\cite{masugate}
& $\checkmark$ & $\checkmark$ & -- \\
Use-time approval binding & Loopjacking, BSC-R~\cite{loopjacking,bscr}
& $\checkmark$ & $\checkmark$ & -- \\
Single-witness contracts & CONTINUITY, AID-Guard, ADR 0030~\cite{continuity,aidguard,agtadr0030}
& $\checkmark$ & $\checkmark$ & $\circ$ \\
Approved-object integrity & Consent Integrity, ReSA~\cite{consentintegrity,resa}
& -- & $\checkmark$ & -- \\
Control-primitive barriers & Stop Means Stop~\cite{stopmeansstop}
& $\circ$ & $\circ$ & -- \\
\midrule
\textbf{Commit-time authorization (\CTA)} & \textbf{\textsc{CommitGate} + atomic sink}
& $\checkmark$ & $\checkmark$ & $\checkmark$ \\
\bottomrule
\end{tabular}
\end{table*}

\noindent\textbf{Information and recovery.}
Notices and consulted status tools helped agents follow the authorization
rule. Revocation can follow their last observation, so the service must
still check authority at commit. Recovery requires an authorized route;
terminal grant or eligibility changes require stopping. Reporting
authorization beside utility exposes unauthorized success and the cost of
safe refusal.

Qwen and Gemma often reached execution limits before OpenHands publication,
including in clean controls. An absent publication cannot establish that an
agent recognized revocation. Model size and family are not isolated from tool format, context
settings, or serving configuration. Fable's provider refusals varied by task
and usually preceded preparation, including in clean runs. Its answered
trajectories are therefore a selected subset. Assigned recovery cohorts
retain refusals, and all cohorts retain effects applied before a later
refusal.

\noindent\textbf{Policy and implementation.}
Omitted dependencies can admit effects outside the intended policy;
unrelated requirements can reject valid ones
(Assumption~\ref{asm:evidence-integrity}). Cedar reproduced all 30
synthetic decisions under our policies and candidate boundaries
(\S\ref{app:policy-evaluation}); host code must still resolve claims and
the sink must enforce them.

\noindent\textbf{Evidence and scope.}
The advisory, retail, scripted, and transport studies estimate no general
failure rate. Thirteen of 17 advisories concern one product, the clause
assignments are ours, and pooling appears only in constructed cases.
Neither scripted repetitions nor model executions with different seeds are
independent samples; no model is fully deterministic at our settings.
The controlled response name \texttt{safe\_commit} may influence decisions.
The service study runs each of 18 synthetic tasks once per condition;
OpenHands does not isolate delegation. OpenHands and retail use our
authorization policies. Service labels come from owner state, but sink and
oracle can share policy or serialization errors.

Every cross-service panel revocation precedes reservation issuance.
Revocation during a reservation's lifetime is tested only at deterministic
logical ticks. The sink checks version before sampling expiry, so a version
abort does not establish whether the reservation was still live; physical
commit can finish just after expiry. Cross-zone timing, sustained contention,
denial of service, issuer outages, and distributed recovery remain untested.

\section{Related Work}
\label{sec:related-work}

\noindent\textbf{Authorization contracts.}
Stale-safe properties bound attribute staleness~\cite{stalesafe}; UCON
re-evaluates rights during use~\cite{uconabc}. Proof-carrying
authentication~\cite{proofcarryingauth} and recency
logics~\cite{stubblebine95,stubblebinewright96} check authority proofs and
credential recency at decision time; consumable credentials become
unavailable everywhere once used~\cite{consumablecredentials}. Cedar~\cite{cedar2024} and
Progent~\cite{progent} allow a request only when one rule holds in full and
no deny rule matches. Request-bound grants, proofs, and capabilities bind
authority to operations~\cite{sudp,caplease,secureclaw,cxi}.
Table~\ref{tab:related-contracts} compares their contracts.

\noindent\textbf{Closest contracts.}
Zhang and Zhang~\cite{authorizationcontinuity} require one live lease,
scoped by resource and effect type, to cover the full effect set; leases
cannot be combined and a surviving alternative suffices. \CTA{} adds
premise, dependency, and eligibility claims from their owners, coherent within
that witness (Proposition~\ref{prop:versioned-grants}). CONTINUITY rechecks revocation
of an exact-action permit at the sink, after admission and within its
lifetime~\cite{continuity}; its unmodified implementation matches the tested contract cases (\S\ref{app:closest-contracts}). ADR~0030 atomically checks and
consumes a one-time, action-bound approval~\cite{agtadr0030}. AID-Guard
requires provider enforcement of an approved version, ETag, or lock
precondition, and rejects root revocation before its authorization
transaction; its atomic profile joins that transaction with the provider
effect~\cite{aidguard}. BSC-R binds revocation, context root, and execution
subject at commit~\cite{bscr}. These systems check one witness or lineage
per action; \CTA{} also selects among complete candidates without pooling
their owners' claims. BSC-R cites \CTA{} and reports 44.531\% invalid commits for its third-party
``CommitGuard-faithful'' reimplementation and 0\% for CONTINUITY on the
latter's attack suite. That projection uses a different policy and sink;
the suite includes additional provenance and transformation
invariants~\cite{bscr}.

\noindent\textbf{Approvals and commit boundaries in agents.}
Loopjacking binds a human approval to the operation evaluated at use and
checks it there~\cite{loopjacking}; Consent Integrity binds the approval
dialog to the exact action~\cite{consentintegrity}; ReSA analyzes approval
consumers that reconstruct the approved object~\cite{resa}; closure-bound
approval covers transitive effects of an approved
command~\cite{approvallaundering}; and Stop Means Stop measures approval
gates and cancellation~\cite{stopmeansstop}.

\noindent\textbf{Coordination and the final effect.}
Atomix~\cite{atomix} coordinates settlement with application-supplied semantic
validation, and Cordon~\cite{cordon} validates lineage, authority, and
approval in task validation. PSS~\cite{masugate} coordinates policy state and
effects through provider transactions or reservations; as on our reservation
path, a transferred grant may outlive source revocation. Lilienthal and
Hong~\cite{mindthegap} and Jiang et~al.~\cite{atomicityagents} address
changing resource or page state, which our approval-only cases hold fixed; a
binding mismatch violates \CTA{} with no race. Wu et~al.~\cite{authorizationexecutiongap} study authority--execution gaps,
Mohammadi et~al.~\cite{ghosttoolcalls} effects visible at dispatch, and agent benchmarks evaluate outputs, traces, or application
state~\cite{webarena,appworld,agentdojo,taubench}.

\section{Conclusion}
\label{sec:conclusion}

An agent can complete a task through an effect its approval never covered
or no longer authorizes, as fixed advisories document. After silent owner
changes, all five models mostly completed service tasks through unauthorized
effects, with Fable measured over answered requests. Notices helped agents
follow the rule; protected refusals could lead to a separately authorized
route.

\CTA{} requires one complete, current justification for the exact effect at
commit. \textsc{CommitGate} authenticates and selects that justification;
the sink enforces it atomically. Protected paths enforced the declared policy
and preserved valid alternatives in the evaluated LangChain, OpenHands,
service, and Atomix cases. Serial admission added at most 0.63~ms in mean
p50 over direct authorization. Machine-checked proofs establish whole-candidate selection,
trace safety for mediated atomic sinks, and exact single-use transfer in the
stated model. Agent systems should establish authority at commit and report
authorization beside task success.

\bibliographystyle{IEEEtran}
\bibliography{references}
\appendices
\section{Ethics Considerations}
This work examines whether an agent's actions remain authorized when they
execute. The experiments use synthetic tasks and data, public software
releases, and isolated local or cloud environments under our control.
Hosted model APIs are used for inference. We conducted no experiments on
human subjects and did not target other users' accounts or production
deployments.

The main dual-use risk is that the examples could help identify similar
authorization failures in deployed agents. We limit the public artifact to
the experiments, records, and defensive implementations needed to reproduce
the released evaluation. Private reports, credentials, and identifying
deployment details are excluded. We privately reported the authorization
issues identified during this work to the relevant maintainers before
submission. The reports remain private.

\section{Open Science}
{\raggedright
The artifact is available at:
\url{https://anonymous.4open.science/r/commit-time-authorization/}.\par}
\texttt{artifact\_manifest.json} indexes specifications, records,
\textsc{CommitGate}, Lean, related work, and tests; \texttt{RESULTS.md} maps
claims to records. With Python~3.10+ and the pinned Lean toolchain,
\texttt{make reproduce-full} rebuilds counts, reruns deterministic services,
checks model records, and builds proofs; \texttt{make reproduce-all} adds
LangGraph replays. With pinned Python~3.13+ environments,
\texttt{make reproduce-sp27} also checks OpenHands, Atomix, retail,
LangChain, and operational studies. Separate commands rerun Atomix, test
recorded Git commits, and fetch pinned CONTINUITY for runtime checks; its
clone is excluded. The OpenHands package records its macOS arm64 SDK
environment. Records include all reported executions, operational warmups and offers,
and ids and source hashes for 270 primary executions and 262 commits.
Inspected WebArena, AgentDojo,
$\tau$-bench, and AppWorld evaluators are pinned. OpenHands records include
model conversations, SDK events, service decisions, and Git objects. Public
projections remove identity and service tokens; non-secret configuration
retains models, settings, retries, prompts, and dependencies. Findings
reproduce from records without provider access.

\section{Machine-Checked Model}
\label{app:boundary-criterion}

Lean~4 checks the model and boundary claims of
\S\ref{sec:model}--\S\ref{sec:defenses}. Structures represent $\sigma$, $c$,
$\Pi$, and $J$; predicates implement $\mathsf{Valid}$, $\mathsf{CTA}$,
$\mathsf{TDiv}$, and $\mathsf{LostCTA}$, with rejection exactly when \CTA{}
fails. The model includes sink transitions, forwarding and selecting traces,
two-state leases with tied issuance, the grant ledger, and cross-service
cases. Induction proves selection over candidates and safety over events;
single use follows from a growing nonce record. The issued-trace part of
Theorem~\ref{thm:trace-safety} requires every logged candidate to have been
presented, hence issued. Binding the version advanced by a write prevents
immediate repetition.

\noindent\textbf{Statements and audit.}
The inventory lists all 139 project theorems: 31 paper statements with
exact elaborated types recorded in the artifact, including
shared-state and single-certificate counterparts of
Theorems~\ref{thm:reservation-transfer} and~\ref{thm:trace-safety}; 20
counterexamples that mark the limit of a check or hypothesis; and 88
supporting lemmas. The audit checks proof completeness and matches Lean's
axiom report, only \texttt{propext}, to the inventory. Lean checks labels
from 60 recorded Qwen executions, one per case, defense, and repetition,
with one outcome in each of 20 cells, against the cross-service shared-state
reservation model. These executions do not instantiate the two-state model
of Theorem~\ref{thm:reservation-transfer}.

\noindent\textbf{Limits of the model.}
Issuer honesty, unique identifiers, and claim provenance are hypotheses
(\S\ref{sec:model}), and Lean does not show that an approval service records
the ledger correctly.
Since $\mathsf{AuthV}_L$ compares version numbers, it presumes that no
resource returns to an earlier one.
Reservation time uses inclusive discrete ticks, while operational consumption
uses an exclusive nanosecond bound at its locked clock sample; no clock
refinement or physical-persistence deadline is proved. Lean does not prove
semantic provenance, cryptographic implementations, agent intent, issue-time
confidentiality, distributed rollback, or crash recovery.

\section{Evaluation Details}
\label{app:task-library}
\label{app:additional-evidence}

\noindent\textbf{Controlled tasks.}
The 54 controlled tasks span three families of six categories with three
tasks each, chosen beforehand to vary how evidence for
authority reaches an effect: browser tasks draw on WebArena, WorkArena, and
OSWorld, tool/state tasks on AppWorld, $\tau$-bench, and OpenHands, and
multi-agent tasks cover routing, memory, verification, review, policy, and
speculative branches. They are synthetic analogues, not a prevalence sample.

\begin{table}[t]
\centering
\caption{Service-agent outcomes, unprotected arm with an owner change:
unauthorized effect with task completed (UC) or not (UN), authorized
completion (A), neither (W), of 36 trajectories (Fable: 22, 23, 27
answered). Untriggered changes (Qwen 2, 2, 6; Gemma 2, 2, 0) count as W.
Brackets: Wilson 95\% interval of the UC share (\%), trajectories taken as
independent.}
\label{tab:service-agent-full}
\footnotesize
\setlength{\tabcolsep}{3pt}
\begin{tabular}{@{}lccc@{}}
\toprule
Model & Silent & Notice & Status \\
\midrule
Qwen & 26/1/0/9 [56, 84] & 15/1/2/18 [27, 58] & 26/0/1/9 [56, 84] \\
Gemma & 34/0/0/2 [82, 98] & 10/0/6/20 [16, 44] & 32/1/0/3 [75, 96] \\
GPT & 25/1/3/7 [53, 82] & 1/0/11/24 [0, 14] & 2/0/9/25 [2, 18] \\
Fable & 17/1/4/0 [57, 90] & 0/0/10/13 [0, 14] & 17/1/3/6 [44, 78] \\
Sonnet & 30/2/2/2 [68, 92] & 0/0/8/28 [0, 10] & 22/2/2/10 [45, 75] \\
\bottomrule
\end{tabular}
\end{table}

\noindent\textbf{Advisory fixes.}
In 14 of the fixes in Table~\ref{tab:advisory-fixes}, the fact checked is
the grant's currency or scope or a field of the executed request, which every \CTA{} policy requires: \CTA{} authorizes the exact
effect, the full input of the durable operation, and how coarse a policy's
grants are is part of what the policy declares. In three, it is a premise
that the policy must declare. In seven with B in
Table~\ref{tab:agent-advisories}, the fix widens what a persisted decision
records or the approver sees, placing the fact in the grant's scope;
\CTA{} requires both. In wj55, the fix checks the policy when the gateway
delivers the queued action; the durable operation is its later execution on
the node. An LLM agent independently coded twelve advisories and the
inclusion of all 21 from the rules and sources alone; it agreed with our
codes on 11 of 12 slots (binding rather than grant scope for the 24mc fix), all 12 requirement
codes, and 18 of 21 inclusion decisions.

\begin{table}[t]
\centering
\caption{What each fix in Table~\ref{tab:agent-advisories} checks and the
\CTA{} claim that expresses it; G grant, P premise, B binding. R: every
\CTA{} policy for the effect requires the fact, with $c$ the full input of
the durable operation; D: the policy must declare it; $\approx$: another
fact or an earlier step.}
\label{tab:advisory-fixes}
\footnotesize
\setlength{\tabcolsep}{2.5pt}
\begin{tabular}{@{}ll>{\raggedright\arraybackslash}p{0.335\columnwidth}>{\raggedright\arraybackslash}p{0.28\columnwidth}c@{}}
\toprule
Id & Fixed & Fix checks & \CTA{} claim & Req. \\
\midrule
24mc & 1.3 & new approval when the entry changes & G scope: command and arguments ($\approx$) & R \\
8g75 & 2026.3.8 & script matches the approval snapshot & P: script digest & D \\
q399 & 2026.3.1 & \texttt{PATH} token resolves to the approved binary & P: \texttt{PATH} resolution & D \\
mwcg & 2026.2.25 & working directory is not a symlink & P: cwd target ($\approx$) & D \\
wj55 & 2026.3.22 & policy at delivery allows the queued action & G currency: policy at execution ($\approx$) & R \\
r22h & 1.3 & install dialog shows the arguments & G scope: arguments shown & R \\
h3f9 & 2026.2.14 & executed argv matches the evaluated command & B: argv & R \\
hwpq & 2026.2.25 & approval matches the exact executed argv & B: argv & R \\
74gc & 2026.8.1 & Allow Always grant keeps the arguments & G scope: arguments & R \\
6j27 & 2026.2.22 & grant keeps the inner executable & G scope: inner executable & R \\
3mq7 & 2026.8.1 & reusable approval bound to the cwd & G scope: cwd & R \\
7jfq & 2026.8.1 & standing grant bound to its node & G scope: target node & R \\
6x2m & 2026.2.23 & approving node matches invoking node & B: target node & R \\
6rcp & 2026.2.24 & approval text shows the full argv & G scope: argv shown & R \\
hjvp & 2026.2.26 & approval bound to argv, cwd, env digest & B: env, with argv and cwd & R \\
pq7j & 2.1.179 & checked-out commit equals the pinned hash & B: commit hash & R \\
q5hj & 2.1.84 & trust lookup validates the worktree link & B: trusted folder ($\approx$) & R \\
\midrule
\multicolumn{5}{@{}p{\columnwidth}@{}}{\emph{Not counted, contestable (4):} group undetermined; no durable effect named (I4, I5); no approver named (I3); not a tool effect (I5).} \\
\bottomrule
\end{tabular}
\end{table}

\noindent\textbf{Service-agent study.}
Of the 18 task ideas mapped to durable writes, one that cannot run safely is
replaced by the nearest write on that asset. Owners keep typed records of
approvals, resource versions, readiness, page-element bindings, and run or
branch state; utility comes from the final state alone, and the agent
makes at most 16 model requests.
Table~\ref{tab:service-agent-full} gives every outcome of the unprotected
arm. A post hoc audit identifies 42 recovery opportunities from the specified
scenario types, retaining untriggered changes and provider refusals.
Fable's 29 answered cases all recover; its 13 refusals include nine before
and four after the change. Across all 108 protected change cases, 63 are
answered and 29 complete.
Under the pair rule of \S\ref{sec:agent-tasks}, 22 (Qwen), 32 (Gemma),
20 (GPT), 19 (Fable), and 34 (Sonnet) of the 36 planned pairs qualify. GPT's
16 excluded pairs differ before the change, 15 in their tool calls; Qwen's
exclusions include trajectories that wrote with the preparation, and 16 of
Fable's 17 contain a refused trajectory. Without the replaced task, Qwen keeps
20 pairs, with unauthorized completions falling from 13 to 3.

\noindent\textbf{Provider refusals.}
Fable's provider refuses every trajectory of three service tasks and none
of six others. Clean and changed trajectories have similar refusal counts
(26/72 and 81/216). Of 107 refusals, 65 occur at the first request; 90 precede
the preparation trigger, 16 follow an applied change, and one follows
preparation in a clean control. Seven follow a protected write refusal at
preflight (two at the next request), yielding 21 refusals in the protected
silent condition against 14 unprotected. Observed unauthorized completions
over all 36 prescribed unprotected cases are 18, 0, and 17 (silent, notice,
status), retaining one commit before a later provider refusal beyond the
answered-only counts in Table~\ref{tab:service-agent}. Controlled refusals
exclude 87 flagged executions, 21 controls, and the one perturbed execution
without a flag; the single-service panel's 42 refusals all concern the
multi-agent surface, leaving 48 answered executions mostly on tool/API.

\noindent\textbf{Model diagnostics.}
In LangChain, Sonnet's first two diagnostics were inconclusive because its
provider refused a read;
a third, using a ticket read, passed. No Sonnet study trajectory was refused.

\noindent\textbf{OpenHands task cohort.}
Fixed offline tests fail on each base revision and pass on its published fix;
production publication binds the exact commit and expected target revision.
Each model runs eight direct and eight delegated trajectories. The
48 of GPT, Fable, and Sonnet contain 51 publication events: 12 authorized
production, 27 authorized staging, and 12 unauthorized production; four
unsafe trajectories also publish to staging later, and
Table~\ref{tab:openhands-tasks} reports their unauthorized outcome. All these
conversations finish, and fixed tests verify candidate code and applied
commits.

\begin{table}[t]
\centering
\caption{Controlled study with Qwen, by family. Unauthorized commits follow
the environment's invalidation flag, which clean controls never carry.}
\label{tab:primary-result}
\footnotesize
\setlength{\tabcolsep}{2pt}
\begin{tabular}{@{}lccccc@{}}
\toprule
Family & \shortstack{Endpoint\\success} &
\shortstack{Authorized\\completion} & \shortstack{Unauthorized\\commit} &
\shortstack{Safe non-\\completion} & \shortstack{Infra.\\failure} \\
\midrule
Browser & 89/90 & 19/90 & 70/90 & 1/90 & 0/90 \\
Tool/state & 83/90 & 18/90 & 65/90 & 7/90 & 0/90 \\
Multi-agent & 90/90 & 18/90 & 72/90 & 0/90 & 0/90 \\
\midrule
All & 262/270 & 55/270 & 207/270 & 8/270 & 0/270 \\
\bottomrule
\end{tabular}
\end{table}

\noindent\textbf{Execution sets.}
Each model has 90 LangGraph executions through ticket and branch sinks,
18 per defense. Every defense preserves six authorized effects. Unguarded
and resource-recheck paths apply twelve and nine unauthorized effects;
the three atomic paths apply none. These results hold for Qwen, Gemma, GPT,
and Sonnet, and cell by cell for Fable's 48 answered executions.
By invalidated fact, Qwen's 207 unauthorized controlled commits
(Table~\ref{tab:primary-result}) split into stale grants or premises (105),
readiness (51), binding (18), and eligibility (33), and the 12 of GPT and
Fable into stale grants or premises (4), readiness (6), and binding (2). No
model output in the cross-service panel is repaired or excluded, and
replaying trace-derived certificates through \textsc{CommitGate} checks
compatibility, not an independent decision.

\subsection{Admission and Policy Evaluation}
\label{app:policy-evaluation}

\begin{table}[t]
\centering
\caption{Single desktop reservation run, in ms: 200 serial operations per arm,
10-s reservation, direct arm first; admission from six later order-balanced
repetitions, so the stages do not sum to the complete path.}
\label{tab:reservation-single-run}
\footnotesize
\setlength{\tabcolsep}{4pt}
\renewcommand{\arraystretch}{1.1}
\begin{tabular}{@{}lrrrr@{}}
\toprule
& \multicolumn{2}{c}{Direct authorization} & \multicolumn{2}{c}{\textsc{CommitGate}} \\
\cmidrule(lr){2-3}\cmidrule(l){4-5}
Stage & p50 & p95 & p50 & p95 \\
\midrule
Admission & 4.83 & 5.67 & 5.09 & 5.81 \\
Issuance & 25.81 & 35.42 & 29.72 & 84.37 \\
Consumption & 29.28 & 44.90 & 32.60 & 82.68 \\
Complete path & 62.02 & 89.94 & 73.01 & 171.17 \\
\bottomrule
\end{tabular}
\end{table}

\noindent\textbf{Reservation workloads.}
Desktop services use an Apple M1 Pro, macOS, Python~3.14.7, SQLite~3.53.4,
and file-backed WAL databases with synchronous FULL. Three
\texttt{n2-standard-4} VMs (4 vCPU, 16~GB) host client, approval, and
resource services in one zone with the same versions and chrony; the control
collocates services on one VM. Serial runs balance arm and topology order
over three blocks per topology. Disjoint contention runs 100 measured bursts
of sixteen processes per cell. The shared-target extension balances arm
order, topology, and window rotation in six blocks in another zone, with its
own collocated control. Desktop contention yields 11,524 completions, 15,090
expiries, and 7,786 version conflicts. Table~\ref{tab:reservation-single-run}
divides its single run by stage. Two VMs add 2.37, 2.98, and 2.35~ms to
\textsc{CommitGate}'s serial p50 at 10~s, 1~s, and 100~ms.

\noindent\textbf{Local admission microbenchmark.}
Over 500 decisions per configuration, median verification grows from 15.417
to 265.105\,$\mu$s for flat certificates with three to 66 claims and from
15.875 to 266.750\,$\mu$s for chained ones, with 2,123 to 3,435 bytes of
traced allocation over 50 decisions, excluding HTTP, storage, state
resolution, and contention.

\noindent\textbf{Mechanism versions.}
Published code runs unchanged: LangChain~1.4.4's approval middleware,
OpenHands SDK~1.47.0 confirmation, LangGraph~1.2.10's \texttt{interrupt},
Cedar~4.13.0, CONTINUITY, and Atomix at \texttt{61e411be}.
The Jenkins vendor regression admits
a rebuild after withdrawal in workflow-cps \texttt{3990.vd281dd77a\_388}
and refuses it in \texttt{3993.v3e20a\_37282f8}. Both current and restored
approval controls succeed. Production source is unchanged.

\noindent\textbf{Encoding \CTA{} in Cedar.}
One Cedar~4.13.0 policy~\cite{cedar2024} requires every declared slot,
trusted issuers, and exact-effect binding within one candidate; host code
checks signatures, resolves currentness, and selects the first permitted
candidate. All 30 synthetic cases match \textsc{CommitGate}, including fresh
alternatives and a review-slot extension. No sink runs: atomic enforcement
and integration effort are unmeasured.

\subsection{Supplementary Checks}
\label{app:supplementary-checks}

\noindent\textbf{LangChain approval path.}
The eight unprotected unauthorized writes follow withdrawal (three cases),
policy change, ticket change (two), an unapproved tool targeting another
ticket, or direct dispatch. The adapter refuses six attempts, including two
pooled attempts after one retry; the sink refuses post-check withdrawal,
the unapproved tool's binding mismatch, and direct dispatch.

LangGraph runs ten framework cases, three sink cases, and two concurrency
arms: withdrawal outside the graph leaves an approved append executable;
the sink refuses it, while an earlier checkpoint asks again. It refuses
reuse under another certificate id and permits one concurrent presentation.
With the lock, withdrawal between check and append waits; without it, append
follows withdrawal.

\noindent\textbf{Transport checks.}
In 24 loopback HTTP executions of clean, invalidating, and post-preflight
cases for ticket and branch services, unguarded execution leaves four
unauthorized commits, resource recheck three, and native conditional write
and \textsc{CommitGate} none. In a two-path Playwright check through system
Chrome, unguarded execution changes the retargeted account while displaying
``Change saved,'' and the protected path withholds the effect.

\begin{table*}[t]
\centering
\caption{Commit-time obligations and supporting cases.}
\label{tab:contract-case-map}
\footnotesize
\setlength{\tabcolsep}{3pt}
\begin{tabular}{@{}>{\raggedright\arraybackslash}p{0.26\textwidth}>{\raggedright\arraybackslash}p{0.43\textwidth}>{\raggedright\arraybackslash}p{0.25\textwidth}@{}}
\toprule
Case & Required decision & Evidence \\
\midrule
Revoked approval, unchanged resource & Reject: resource equality does not supply a current grant & Cross-service panel \\
Stale A, complete current B & Accept B: divergence of A does not remove all authorization & Service and Atomix panels \\
Missing required evidence & Withhold: no complete candidate satisfies the declared policy & Service panel; schema-extension test \\
Revocation after positive preflight & Reject at the atomic sink: earlier admission remains provisional & Service and Atomix panels \\
\bottomrule
\end{tabular}
\end{table*}

\noindent\textbf{Deterministic studies.}
Each deterministic case runs once, without a model. Binding cases present a
complete sibling-effect candidate or another effect's grant; partial cases
present grant-only and binding-only candidates, and positive controls a
complete candidate. A \emph{merged} candidate joins partial candidates for
an authorized effect; a \emph{pooled} one joins current claims from two
invalidated candidates. Epoch-only grants admit foreign grants; omitting
candidate names admits merged and pooled candidates. Signing the effect id
and payload digest into grants refuses all six foreign-grant candidates,
four with the requested target and operation. Signing candidate identity
(22 adapter, 7 evaluator, 59 reservation-path lines) refuses merged and
pooled candidates, which unguarded and resource-recheck paths apply; the
approval service refuses reservation issuance from a merged candidate.
Cross-service lookup finds no candidate for ticket~42 in the binding case,
so preflight refuses it. A reservation issued at tick~10 expires at~12;
source revocation at~11 still permits consumption through~12, while tick~13
and repeat presentation are refused.

The specificity panel crosses three systems, four defenses, two claim
scopes, one clean, two invalidating, and three benign cases: superseding another
effect's approval, one of two alternatives, or changing unrelated resource
state. Both protected paths complete all nine benign cases with effect-scoped
claims and none with service-wide counters, refusing all six invalidating
cases in both scopes. Whole-resource \texttt{If-Match} loses the
ticket-comment case; this panel includes no other contract implementation.

\section{Adapter and Sink Contracts}
\label{app:independent-oracle}
\label{app:adapter-contract}

\begin{table}[t]
\centering
\caption{Cases executed on all three systems under all four defenses.}
\label{tab:independent-cases}
\footnotesize
\setlength{\tabcolsep}{2.5pt}
\begin{tabular}{@{}lll@{}}
\toprule
Case & Status & Property exercised \\
\midrule
Clean & valid & clean completion \\
Benign concurrent & valid & a counter no check reads changes \\
Alternative grant & valid & one current justification suffices \\
Stale unrelated & valid & the same unread counter changes \\
Invalidating & invalid & target, version, approval, or path drift \\
Post-preflight race & invalid & atomic sink coupling \\
Forged claim & invalid & issuer authentication \\
Missing declared slot & invalid & required evidence is absent \\
\bottomrule
\end{tabular}
\end{table}

\begin{table}[t]
\centering
\caption{Closest published contracts. R: refused; A: accepted; App:
application responsibility; --: not stated. CONTINUITY verdicts are also
executed; letters refer to \S\ref{app:closest-contracts}.}
\label{tab:closest-contracts}
\footnotesize
\setlength{\tabcolsep}{2.3pt}
\begin{tabular}{@{}lcccccc@{}}
\toprule
Case & \cite{authorizationcontinuity} & \cite{continuity} & \cite{agtadr0030} & \cite{aidguard}$^g$ & \cite{bscr} & \CTA \\
\midrule
Grant revoked; resource fixed & R & R & R & R & R & R \\
Grant for another effect & R$^a$ & R & R & R & R & R \\
Live grant; premise changed & --$^b$ & A$^c$ & App & R & -- & R \\
Pooled candidate & --$^b$ & --$^e$ & -- & -- & -- & R \\
Surviving alternative & A & A$^f$ & -- & -- & -- & A \\
Post-check revocation & -- & R & R & R & R & R \\
Source revoked in reservation & -- & R$^c$ & -- & -- & -- & A$^d$ \\
\bottomrule
\end{tabular}
\end{table}

\noindent\textbf{Authoritative systems.}
The deterministic services label effects independently of model output:
the browser records the mutated account and document epoch despite selector
retargeting; the ticket sink resolves version and approval; the branch sink
checks the completed worker's publication capability.
Table~\ref{tab:independent-cases} lists cases;
Table~\ref{tab:contract-case-map} maps them to obligations. Defenses start
from identical state and evidence; predeclared labels are never supplied to
sinks. All three preflights pass when the monitor's resolver falsely reports
current evidence; the oracle labels every applied effect unauthorized.

\noindent\textbf{Integrating a protected effect.}
\label{sec:deployment-requirements}
The adapter author fixes $c$ and its atomic commit operation
(Table~\ref{tab:policy-manifest}), then assigns each slot a type, authoritative
issuer, and resolver from tool schemas, approval manifests, or orchestrator
state. Missing slots invalidate candidates, including after policy extension.
Omitted dependencies escape enforcement; needless slots cause benign aborts.
The negative control detects verdict reuse; shared-policy errors remain
undetected. Tests check derived-grant replay, expiry, wrong effects, and
validation atomic with mutation during races.

\section{Closest Published Contracts}
\label{app:closest-contracts}

All 15 CONTINUITY cases, including four controls, match predeclared outcomes
at unmodified commit \texttt{59cfe609} (Table~\ref{tab:closest-contracts}).
(a)~A lease scopes resources and effect types, without distinguishing
same-type effects on one resource; (b)~leases are not combined but carry no
premises~\cite{authorizationcontinuity}.
(c)~CONTINUITY rechecks source revocation; a version-7 permit executes
on ticket version~8~\cite{continuity}.
(d)~A derived grant remains valid until expiry, at most
$\tau+2\epsilon$ after issuance (Theorem~\ref{thm:reservation-transfer}).
(e)~Root-grant binding prevents object mixing across candidates; approval~3's own
version-7 bundle executes at version~8.
(f)~The caller selects one permit per call; the live permit commits once
after the stale permit's refusal.
(g)~AID-Guard's atomic provider profile; submitted-attempt authorization
linearizes before the provider effect~\cite{aidguard}.
Applications supply oracles and complete views~\cite{authorizationcontinuity},
mediation and provider transactions~\cite{continuity}, and facts outside
approval binding~\cite{agtadr0030}.

\end{document}